\documentclass[english,english,preprint,amsmath,floatfix]{revtex4-1}
\usepackage[T1]{fontenc}
\usepackage[latin9]{inputenc}
\usepackage{color}
\usepackage{babel}
\usepackage{float}
\usepackage{amstext}
\usepackage{graphicx}
\usepackage[unicode=true,pdfusetitle,
 bookmarks=true,bookmarksnumbered=false,bookmarksopen=false,
 breaklinks=false,pdfborder={0 0 1},backref=false,colorlinks=true]
 {hyperref}

\makeatletter

\providecommand{\tabularnewline}{\\}

\@ifundefined{textcolor}{}
{%
 \definecolor{BLACK}{gray}{0}
 \definecolor{WHITE}{gray}{1}
 \definecolor{RED}{rgb}{1,0,0}
 \definecolor{GREEN}{rgb}{0,1,0}
 \definecolor{BLUE}{rgb}{0,0,1}
 \definecolor{CYAN}{cmyk}{1,0,0,0}
 \definecolor{MAGENTA}{cmyk}{0,1,0,0}
 \definecolor{YELLOW}{cmyk}{0,0,1,0}
}

\usepackage{babel}
\usepackage{babel}
\setcitestyle{super}

\makeatother

\begin{document}

\title{{\normalsize{}Theory of Triplet Optical Absorption in Oligoacenes:
From Naphthalene to Heptacene}}

\author{{\normalsize{}Himanshu Chakraborty }}

\author{{\normalsize{}Alok Shukla}}

\affiliation{Department of Physics, Indian Institute of Technology Bombay, Powai,
Mumbai 400076, INDIA}

\email{chakraborty.himanshu@gmail.com, shukla@phy.iitb.ac.in }

\begin{abstract}
In this paper we present a detailed theory of the triplet states of
oligoacenes containing up to seven rings, \emph{i.e}., starting from
naphthalene all the way up to heptacene. In particular, we present
results on the optical absorption from the first triplet excited state
$1^{3}B_{2u}^{+}$ of these oligomers, computed using the Pariser-Parr-Pople
(PPP) model Hamiltonian, and a correlated electron approach employing
the configuration-interaction (CI) methodology at various levels.
Excitation energies of various triplets states obtained by our calculations
are in good agreement with the experimental results, where available.
The computed triplet spectra of oligoacenes exhibits rich structure
dominated by two absorption peaks of high intensities, which are well
separated in energy, and are caused by photons polarized along the
conjugation direction. This prediction of ours can be tested in future
experiments performed on oriented samples of oligoacenes. 
\end{abstract}



\maketitle

\section{{\normalsize{}Introduction}}

Conjugated polymers form a class of materials which are strong candidates
for building the next generation of optoelectronic devices.\cite{conjpol-general,Geerts1998}
In order to be able utilize them for these purposes, a thorough understanding
of their electronic structure and optical properties is essential.
Most of the polymers useful for optoelectronic devices have singlet
ground states, and, therefore, singlet excited states determine their
optical properties. As a result of that, most of the theoretical studies
of optical properties of conjugated polymers, have concentrated on
the absorption spectra in the singlet manifold.\cite{barford-book}
However, the triplet states of these systems have become important
because of the possibility of ``singlet fission'', \emph{i.e.},
a singlet excited state decaying into two triplets leading to higher
photo voltaic yield.\cite{singlet-fission-michl,singlet-fission-michl2}
Furthermore, within the tight-binding model, the lowest triplet state
has the same orbital occupancy and spatial symmetry as the lowest
optically active singlet state, thus rendering them degenerate. Therefore,
differences between these two states will be due to electron correlation
effects, and, thus, a study of triplet states provides us with an
insight into the role of electron correlations in that material.\cite{Shukla-PPV}

Recently, polyacenes have been considered as strong candidates for
optoelectronic device applications such as light-emitting diodes,
and field effect transistors.\cite{bendikov_chem_rev_2004,anthony_chem_rev_2006,bjorseth1983handbook,harvey1991polycyclic}
Furthermore, because of their structural similarities to zigzag graphene
nanoribbons,\cite{bendikov_chem_rev_2004,Anthony2008} the research
effort involving various oligoacenes has further intensified. Although,
most of the studies have concentrated on oligoacenes ranging from
naphthalene to pentacene, several recent works have reported the synthesis
of longer oligomers such as hexacene, heptacene, and beyond.\cite{neckers2008,mondal_bettinger_JACS2009,Kaur_non_JACS2010,Tonshoff2010}
Over the years, optical properties of oligoacenes have been studied
extensively, however, most of these studies have been confined to
their ground state absorption into higher singlet states,\cite{bendikov_chem_rev_2004,anthony_chem_rev_2006,sony-acene-lo}
with the number of studies dedicated to triplet states being far fewer.
Triplet states of naphthalene have been probed experimentally by Lewis
\emph{et al}.,\cite{kasha1944} McClure,\cite{mcclure1949} Hunziker,\cite{hunziker1969,hunziker1972}
and Meyer \emph{et al}.\cite{leclercq1971} Similarly, experimental
measurements of the triplet states in longer stable oligomers namely
anthracene,\cite{McGlynn1964,wright1955,Allan,leclercq1971} tetracene,\cite{McGlynn1964,Sabbatini86(1982),wright1955,windsor1958,pavlopoulos1972,leclercq1971}
and pentacene,\cite{Burgos83(1977),pavlopoulos1972,roberge1972} have
also been performed. A few triplet state measurements of relatively
unstable hexacene,\cite{Angliker} and heptacene\cite{neckers2008}
also exist.\textbf{ }Recently, several experimental measurements of
the triplet states have also been performed on thin films, crystalline,
and dimeric samples of tetracene\cite{burdettandbardeen2012,thorsmolleetal2009}
and pentacene,\cite{angeretal2012,raoetal2010,muntwileretal2009,thorsmolleetal2009,marciniaketal2009,jundtetal1995}
predicting that the lowest triplet state is of charge transfer type.\textbf{ }

As far as theoretical studies of the triplet states of oligoacenes
are concerned, using a Pariser-Parr-Pople (PPP) model type semi-empirical
approach,\cite{ppp-pople,ppp-pariser-parr} early calculations were
performed by Pariser.\cite{pariser1956} Subsequently, again using
the PPP model, low-order configuration interaction (CI) calculations
of triplet states of various oligoacenes were performed by Groot and
Hoytink,\cite{hoytink1966} and Angliker \emph{et al.}\cite{Angliker}
A self-consistent-field random-phase-approximation (SCF-RPA) scheme
also within the PPP model was employed by Baldo \emph{et al.}\cite{tomasello1982}
to perform calculations of triplet states in polyacenes. Large scale
density matrix renormalization group (DMRG) based calculations using
the PPP model, have been performed by Ramasesha and co-workers,\cite{Ramasesha-acene-2002}
and recently, they reported exact diagonalization (full CI) calculations,
for tetracene.\cite{patiandramasesha2014}Complete neglect of differential
overlap (CNDO) based CI calculations employing CNDO/S2 parameterization
were done by Lipari and Duke,\cite{lipari1975} while \textcolor{black}{CNDO/S-CI
method was used by Sanche and co-workers\cite{sanche1990} for triplet
excited state calculations}. Gao \emph{et al}.\cite{gao2002} employed
a spin Hamiltonian and valence bond approach to compute the triplet
states of oligoacenes. As far as \emph{ab initio} calculations are
concerned, complete-active-space self-consistent field (CASSCF) and
perturbation theory (PT2F) calculations on these systems were performed
by Rubio \emph{et al},\cite{roos1994} while multi-reference Møller-Plesset
(MRMP) theory calculations were reported by Hirao and co-workers,\cite{hirao1996,hirao1999}
and Zimmerman et al.\cite{zimmermanetal2010} Chan and co-workers\cite{chan2007}
reported calculations of triplet states of oligoacenes combining the
CASSCF and density-matrix renormalization group (DMRG) approaches,
while coupled-cluster theory (CCSD(T)) based calculations were performed
by Hajgato \emph{et al}.\cite{hajgatoetal2009} Numerous first-principles
density-functional theory (DFT) based calculations have also been
performed by various authors which include works of Houk \emph{et
al.},\cite{Houk} Quarti \emph{et al}.,\cite{quartietal2011}\textbf{
}Anger and co-workers,\cite{angeretal2012} Hummer \emph{et al}.,\cite{hummeranddraxl2005}
Bendikov \emph{et al}.,\cite{bendikov_chem_rev_2004} Nguyen \emph{et
al}.,\cite{patcher-dft} and spin-polarized DFT calculations by Jiang
and Dai.\cite{jiang2008} Most of the theoretical studies mentioned
above either concentrated on a class of triplet excited states of
small oligomers, or only on the energetics of their lowest triplet
excited state. However, none of the earlier works have reported the
calculations of triplet optical absorption spectra, which involves
not only the energetics and the wave functions of a number of triplet
excited states, but also computation of their transition dipole moments
with respect to the lowest triplet state.

In this work, we have performed large-scale correlated electron calculations
of optical absorption in the triplet manifolds of oligoacenes ranging
from naphthalene up to heptacene, using the multi-reference singles-doubles
CI (MRSDCI) method, employing the PPP model Hamiltonian. The detailed
results obtained for the low-lying triplet excited states of these
acenes are presented and compared with different experimental and
other theoretical works. Our calculated energies of the lowest triplet
excited states, $1^{3}B_{2u}^{+}$, $1^{3}B_{1g}^{-}$, $1^{3}B_{1g}^{+}$,
$1^{3}A_{g}^{-},$ and $1^{3}A_{g}^{+}$ for these oligoacenes show
very good agreement with the experimental results. The calculated
triplet absorption spectra of these oligoacenes reveal two long-axis
polarized intense peaks which are well separated in energy, as against
one intense peak observed in their singlet absorption spectra. This
observation is in agreement with our earlier triplet absorption calculations
of the longer acenes namely octacene, nonacene and decacene,\cite{chakrabortyandshukla2013}
and can be tested in future experiments on the oriented samples of
oligoacenes.

Remainder of this paper is organized as follows. In the section \ref{sec:theory}
we describe the theoretical methodology employed for performing these
calculations. Then, in section \ref{sec:results}, we present and
discuss our results. Finally, in section \ref{sec:conclusion}, we
present our concluding remarks, \textcolor{black}{while the convergence
of lowest triplet excitation energies with respect to the PPP parameters,
influence of geometry on the triplet energies, and the convergence
of the MRSDCI excitation energies along with the quantitative details
of the excited states contributing to the absorption spectra, and
their many-particle wave functions, are presented in the Appendixes
A and B.}

\section{{\normalsize{}Theory }}

\label{sec:theory}

\textcolor{black}{Fig.\ref{fig-acene} presents the schematic structure
of an oligocene lying in the $xy$-plane, where the $x$-axis is assumed
to be the conjugation direction. For the purpose of computational
simplicity, a highly symmetric geometry of oligoacenes has been employed
in these calculations, with all nearest-neighbor carbon-carbon bond
lengths fixed at 1.4 \AA , and all the bond angles assumed to be 120$^{o}$.
However, explicit calculations have been performed, in Appendix \ref{sec:appendixA},
to demonstrate that the triplet optical absorption spectra obtained
using realistic asymmetric geometries, with non-uniform bond lengths
and bond angles, do not exhibit any significant differences as compared
to the symmetric geometry adopted in the present work. }

\begin{center}
\textcolor{black}{}
\begin{figure}[H]
\textcolor{black}{\centering{}\includegraphics[width=4cm]{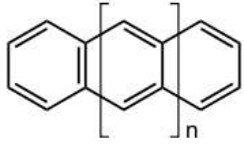}\label{sec:fig-acene}\protect\protect\protect\caption{Schematic drawings of polyacenes\label{fig-acene}}
}
\end{figure}

\par\end{center}

\textcolor{black}{The symmetric structure can be viewed as two identical
polyene chains of suitable lengths, coupled along the $y$-direction.
We have used the PPP model Hamiltonian\cite{ppp-pople,ppp-pariser-parr}
for the correlated calculations, which can be decomposed as}

\textcolor{black}{{} 
\[
H=H_{C_{1}}+H_{C_{2}}+H_{C_{1}C_{2}}+H_{ee},
\]
where the first two terms, $H_{C_{1}}$ and $H_{C_{2}}$ depict the
one-electron Hamiltonians for the carbon atoms located on the upper
and the lower polyene like chains, respectively. The third term, $H_{C_{1}C_{2}}$
is the one-electron hopping between the two chains, and the last term,
H$_{ee}$ represents the electron-electron repulsion. The individual
terms can be written as follows in the second-quantized notation, }

\textcolor{black}{
\[
H_{C_{1}}=-t_{0}\sum_{\langle k,k'\rangle}B_{k,k'},
\]
}

\textcolor{black}{
\[
H_{C_{2}}=-t_{0}\sum_{\langle\mu,\nu\rangle}B_{\mu,\nu},
\]
}

\textcolor{black}{
\[
H_{C_{1}C_{2}}=-t_{\perp}\sum_{\langle k,\mu\rangle}B_{k,\mu}.
\]
}

\textcolor{black}{
\begin{eqnarray*}
H_{ee} & = & U\sum_{i}n_{i\uparrow}n_{i\downarrow}+\frac{1}{2}\sum_{i\neq j}V_{i,j}(n_{i}-1)(n_{j}-1)
\end{eqnarray*}
In the equation above, the carbon atoms on the upper and lower polyene
chains are denoted by $k$, $k'$ and $\mu,\nu$ respectively, whereas
$i$ and $j$ depict all the atoms of the oligomer. The nearest neighbors
are represented by the symbol $\langle...\rangle$, and $B_{i,j}=\sum_{\sigma}(c_{i,\sigma}^{\dagger}c_{j,\sigma}+h.c.)$,
where $h.c.$ denotes the Hermitian conjugate. The one-electron hops
are denoted by the matrix elements $t_{0}$, and $t_{\perp}$. We
took the quantitative value of the hopping matrix elements as $t_{0}=2.4$
eV for both intracell and intercell hoppings, and $t_{\perp}=t_{0}$,
in agreement with the undimerized ground state for polyacene asserted
by Raghu }\textcolor{black}{\emph{et. al}}\textcolor{black}{.\cite{Ramasesha-acene-2002}}

\textcolor{black}{The Coulomb interactions are parametrized according
to the Ohno relationship,\cite{Ohno1964} 
\[
V_{i,j}=U/\kappa_{i,j}(1+0.6117R_{i,j}^{2})^{1/2}\;\mbox{,}
\]
}

\textcolor{black}{where, $\kappa_{i,j}$ denotes the dielectric constant
of the system to reproduce the effects of screening. The on-site repulsion
term is depicted by $U$, while $R_{i,j}$ implies the distance in
Å between the $i$th carbon and the $j$th carbon. The present calculations
have been carried out using: (a) ``standard parameters''\cite{Ohno1964}
with $U=11.13$ eV and $\kappa_{i,j}=1.0$, and (b) ``screened parameters''\cite{Chandross}
with $U=8.0$ eV and $\kappa_{i,j}=2.0$ ($i\neq j$) and $\kappa_{i,i}=1.0$.
Chandross and Mazumdar,\cite{Chandross} devised the screened parameters,
so as to account for the inter-chain screening effects in phenylene
based polymers. However, they can also be seen to describe the effect
of the host in solution or thin-film based experiments. In Appendix
\ref{appa-parameters}, we examine the effect of variations in the
values of various PPP parameters, on the computed excitation energy
of the lowest triplet state of naphthalene. }

\textcolor{black}{Several authors compute high-spin states such as
the triplet state using broken symmetry unrestricted HF (UHF) approach,
or its DFT counterpart unrestricted DFT (UDFT) methods.\cite{Houk,wudl2004,jiang2008}
Although such wave functions have $z$-component of the total spin
$(S_{z})$ as a good quantum number, but they are not eigenstates
of the total spin $(S^{2})$. Such calculations normally lead to results
which can be seen as artifacts of the method because of the large
spin contamination associated with the corresponding many-particle
wave functions.\cite{hajgatoetal2009,deleuze2011,deleuze2012} Therefore,
in this work we use only the restricted HF method (RHF), coupled with
the CI approach where the many-particle wave function is an eigenfunction
of both $S^{2}$ and $S_{z}$ operators, in addition to the corresponding
point group operators. Thus our calculations are initiated by performing
RHF calculations, employing the PPP Hamiltonian, using a computer
code developed in our group.\cite{priya_ppp} All the resultant HF
molecular orbitals are treated as active orbitals. For shorter acenes,
full configuration interaction (FCI) method was used, while for the
longer ones quadruple CI (QCI) and MRSDCI methods were employed. In
particular, FCI method was used for naphthalene and anthracene, while
the QCI/MRSDCI methods were employed from tetracene onwards. The MRSDCI
method,\cite{peyerimhoff_CI,peyerimhoff_energy_CI} is a well known
technique which includes electron-correlation effects beyond the mean-field
both for the ground and excited states of molecular systems. In this
approach, the CI matrix is constructed by generating singly and doubly
excited configurations with respect to a given set of reference configurations
which are specific to the states being targeted in the calculation.
The calculations are performed in an iterative manner until acceptable
convergence has been achieved. We have used this methodology extensively
within the PPP model to study the optical properties of a number of
conjugated polymers,\cite{Shukla-PPV,Shukla-THG,Shukla-TPA,Shukla2,sony-acene-lo,chakrabortyandshukla2013}
and refer the reader to those papers for the technical details associated
with the approach.}

\section{{\normalsize{}Results and Discussions}}

\label{sec:results}

\begin{center}
\begin{table}[H]
\protect\textcolor{black}{\protect\protect\caption{\textcolor{black}{Total number of orbitals of different symmetries
used in our calculations, for oligoacenes (acene-$n)$ of increasing
length, $n$. Both the standard and screened parameters have the same
number of orbitals for the mentioned symmetries. The number in the
parenthesis for each symmetry indicates the number of doubly-occupied
orbitals of that symmetry in the Hartree-Fock ground state. The number
of unoccupied orbitals of a given symmetry can be obtained simply
by subtracting the number of occupied orbitals of that symmetry, from
the total number of orbitals of the symmetry concerned. \label{tab:orbitals-symmetries} }}
}

\centering{}\textcolor{black}{\centering{}}%
\begin{tabular}{ccccc}
\hline 
\textcolor{black}{\ \ $n$\ \ }  & \textcolor{black}{$\ \ A_{g}\ \ $ }  & \textcolor{black}{$\ \ B_{2u}\ \ $ }  & \textcolor{black}{$\ \ B_{3u}\ \ $ }  & \textcolor{black}{$\ \ B_{1g}\ \ $}\tabularnewline
\hline 
\textcolor{black}{$2$ }  & \textcolor{black}{$3$(2) }  & \textcolor{black}{$3$(1) }  & \textcolor{black}{$2(1)$ }  & \textcolor{black}{$2(1)$}\tabularnewline
\textcolor{black}{$3$ }  & \textcolor{black}{$4$(2) }  & \textcolor{black}{$4(2)$ }  & \textcolor{black}{$3(2)$ }  & \textcolor{black}{$3(1)$}\tabularnewline
\textcolor{black}{$4$ }  & \textcolor{black}{$5(3)$ }  & \textcolor{black}{$5(2)$ }  & \textcolor{black}{$4(2)$ }  & \textcolor{black}{$4(2)$}\tabularnewline
\textcolor{black}{$5$ }  & \textcolor{black}{$6$(3) }  & \textcolor{black}{$6(3)$ }  & \textcolor{black}{$5(3)$ }  & \textcolor{black}{$5(2)$}\tabularnewline
\textcolor{black}{$6$ }  & \textcolor{black}{$7(4)$ }  & \textcolor{black}{$7(3)$ }  & \textcolor{black}{$6(3)$ }  & \textcolor{black}{$6(3)$}\tabularnewline
\textcolor{black}{$7$ }  & \textcolor{black}{$8($4) }  & \textcolor{black}{$8(4)$ }  & \textcolor{black}{$7(4)$ }  & \textcolor{black}{$7(3)$}\tabularnewline
\hline 
\end{tabular}
\end{table}

\par\end{center}

\begin{table}[H]
\protect\protect\protect\caption{The sizes of the CI matrix diagonalized, for different symmetries
of oligoacenes (acene-$n)$ of increasing length, $n$. Below std
and scr refer to standard and screened parameters, respectively. \label{tab:convergence}
Superscripts $a,\; b,\ldots$ etc. refer to the type of CI calculations
performed, and are explained below.}

\begin{centering}
\begin{tabular}{cccccc}
\hline 
$n$  & $^{1}A_{g}^{-}$  & $^{1}B_{2u}^{+}$  & $^{3}B_{2u}^{+}$  & $^{3}A{}_{g}^{-}$  & $^{3}B_{1g}^{-}$\tabularnewline
\hline 
$2$  & $4936$$^{a}$  & $4794{}^{a}$  & $7370{}^{a}$  & $7360{}^{a}$  & $7440{}^{a}$\tabularnewline
$3$  & $623576{}^{a}$  & $618478{}^{a}$  & $1099182{}^{a}$  & $1106140{}^{a}$  & $1104544{}^{a}$\tabularnewline
$4$  & $193538^{b}$  & $335325{}^{b}$  & $614865{}^{b}$  & $201879{}^{c}$  & $217203$$^{c}$\tabularnewline
 &  &  &  & $224735{}^{c}$  & $224266$$^{c}$\tabularnewline
$5$  & $1002597^{b}$  & $1707243^{b}$  & $3202299^{b}$  & $582278{}^{c}$  & $490532{}^{c}$\tabularnewline
 &  &  &  & $621397{}^{d}$  & $531551{}^{d}$\tabularnewline
$6$  & $3940254^{b}$  & $6434183^{b}$  & $12234931^{b}$  & $1231948{}^{c}$  & $1156916{}^{c}$\tabularnewline
 &  &  &  & $1443726{}^{d}$  & $1107262{}^{d}$\tabularnewline
$7$  & $12703819^{b}$  & $19663495^{b}$  & $37724739^{b}$  & $2414274{}^{c}$  & $1848466{}^{c}$\tabularnewline
 &  &  &  & $2611198{}^{d}$  & $1795626{}^{d}$\tabularnewline
\hline 
\end{tabular}
\par\end{centering}

\begin{tabular}{c}
$^{a}$FCI\tabularnewline
$^{b}$QCI\tabularnewline
$^{c}$MRSDCI (std)\tabularnewline
$^{d}$MRSDCI (scr)\tabularnewline
\end{tabular}
\end{table}

Before discussing our results, \textcolor{black}{we present the total
number of orbitals used for the different symmetries, in the Table}\textcolor{red}{{}
\ref{tab:orbitals-symmetries},} while in Table \textcolor{black}{\ref{tab:convergence}}
the dimensions of the CI matrices employed for calculating the ground
and excited states of different symmetries, for oligoacenes of increasing
lengths. We note that the largest QCI calculation performed on heptacene
involved dimension in excess of thirty seven million configurations.
Thus, given the large-scale nature of these calculations, we are confident
that electron-correlation effects have been accounted for in an adequate
manner. Because of the large dimensions of these matrices, typically
lowest 50--70 of their eigenroots were computed using Davidson algorithm
as implemented in the MELD program.\cite{meld}In spite of this restriction
in the number of computed eigenvectors, the energy region covered
for computing the absorption spectra was sufficiently large so as
to allow comparison of our results with the experiments in a broad
spectral range. For the calculations presented here, we have ensured
that the convergence with respect to the size of the CI expansion
has been achieved, and in Appendix \ref{sec:appendixA}, for the case
of anthracene, we have demonstrated the convergence of the MRSDCI
calculations by comparing them with the FCI result.

\begin{center}
\begin{figure}[H]
\begin{centering}
\includegraphics[clip,width=8cm]{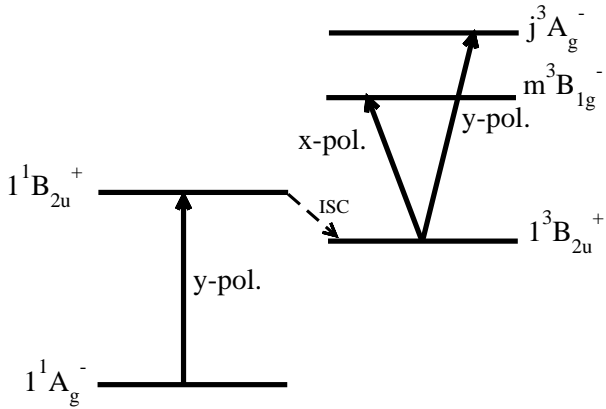} 
\par\end{centering}

\protect\protect\textcolor{black}{\protect\caption{\textcolor{black}{Diagram depicting the important states in the triplet
excited state absorption in oligoacenes, along with their polarization
characteristics. The optical absorption is denoted by the arrows connecting
two states, and the polarization directions are denoted next to them.
Inter-system crossing (ISC) is shown as dotted lines. Location of
the states is not up to scale.\label{fig:PA}}}
}
\end{figure}

\par\end{center}

Triplet excited state absorption can be explained by means of a schematic
diagram shown in Fig. \ref{fig:PA}. Light absorption takes place
from $1^{3}B_{2u}^{+}$ ($T_{1}$) state through long-axis ($x$)
or short-axis ($y$) polarized photons leading the system to $^{3}B_{1g}^{-}$
or $^{3}A_{g}^{-}$ type excited states. As far as the energetics
of various excited states are concerned, our results are summarized
in Tables \ref{tab:Comparison-2-4} and \ref{tab:Comparison-5-7},
where experimental and theoretical results obtained by other authors
are also presented. In triplet absorption experiments, the excitation
energies are measured with respect to the reference $1^{3}B_{2u}^{+}$
state. Therefore, to keep the comparison with the experiments transparent,
in Tables \ref{tab:Comparison-2-4} and \ref{tab:Comparison-5-7}
we present the energies of $^{3}B_{1g}$ or $^{3}A_{g}$ type excited
states with respect to the $1^{3}B_{2u}^{+}$ state. However, excitation
energies of these states with respect to the $1^{1}A_{g}^{-}$ ground
state are presented in various tables in \textcolor{black}{Appendix
\ref{sec:appendixB}, as also in Fig. \ref{fig:Convergence-acene-size}}.

\subsection{Energies of Triplet States}

Before discussing the cases of individual oligomers, first we make
a few general comments about the results presented in Tables \ref{tab:Comparison-2-4}
and \ref{tab:Comparison-5-7}. \textcolor{black}{Additionally, the
plots of the excitation energies of the lowest excited triplet states
of various symmetries, namely, $1^{3}B_{2u}^{+},$ $1^{3}B_{1g}^{+},$$1^{3}A_{g}^{+},$
$1^{3}B_{1g}^{-}$ and $1^{3}A_{g}^{-}$, with respect to the ground
state $1^{1}A_{g}^{-}$, as a function of the increasing acene size
($n$) are presented in Figure \ref{fig:Convergence-acene-size}.
Before we discuss our results in detail, we would like to emphasize
the fact that because PPP model does not take vibrational degrees
of freedoms into account, our calculated values of excitation energies
of various states are essentially ``vertical excitation energies''.
However, experimental measured values of the same quantities, depending
upon the technique employed, may, or may not, be vertical values.
Therefore, while comparing our results with the experimental ones
in the following discussion, we also mention the experimental technique
involved. We also have to bear in mind that one of the reasons behind
the disagreement between the theory and the experiment could be the
fact that the experiment in question may not be measuring the vertical
excitation energies.}

\begin{center}
\begin{figure}[H]
\textcolor{red}{\protect}\textcolor{black}{\protect\protect\caption{\label{fig:Convergence-acene-size}\textcolor{black}{Variation of
the excitation energies (with respect to the ground state,$1^{1}A_{g}^{-}$)
of the lowest excited triplet states of various symmetries, as a function
of acene size ($n$) calculated using: (a) standard parameters, and
(b) screened parameters.}}
}

\centering{}\textcolor{red}{\includegraphics[angle=-90,width=16cm]{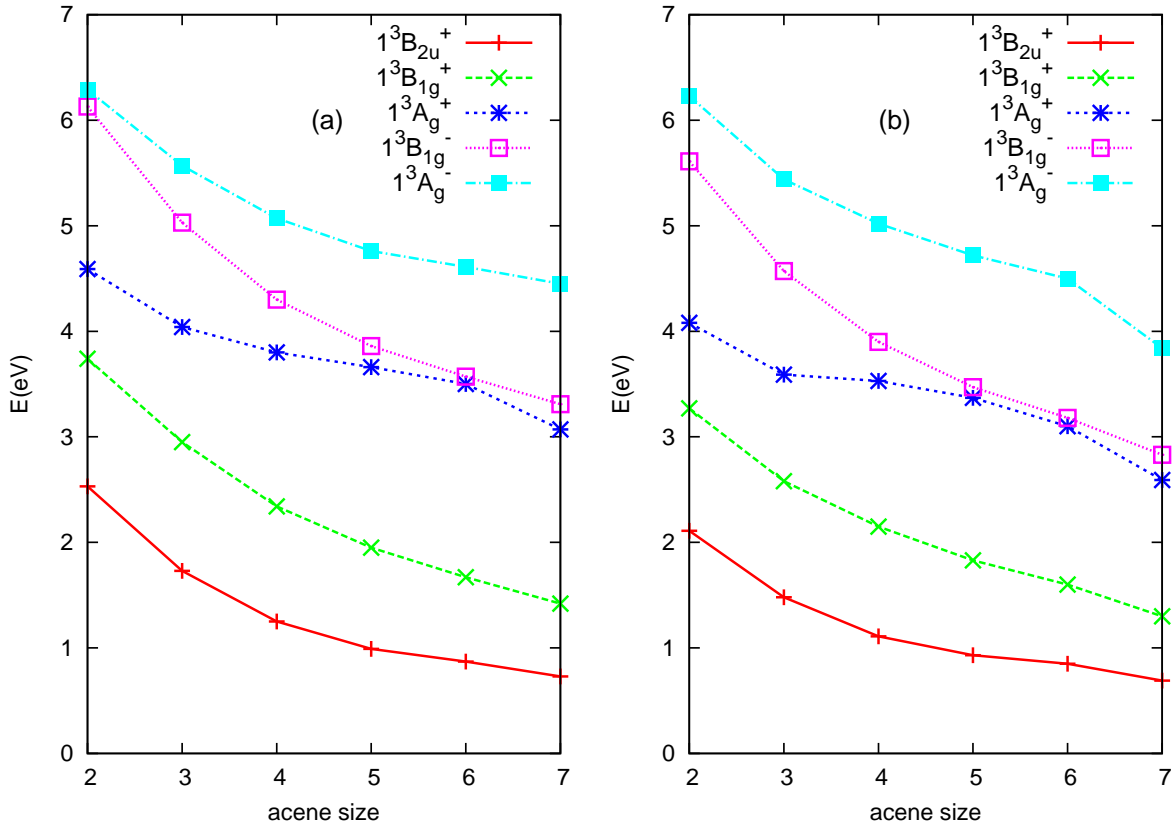}} 
\end{figure}

\par\end{center}

As far as the excitation energy of first triplet excited state $1^{3}B_{2u}^{+}$
state is concerned, our standard parameter based results are in better
agreement with the experiments as compared to the screened parameter
ones, for naphthalene, anthracene and tetracene. From pentacene onwards,
the results obtained with the two parameter sets tend to merge, however,
experimental results are not available for the $1^{3}B_{2u}^{+}$
state of heptacene.

Other important triplet states for oligoacenes are $1^{3}B_{1g}^{+}$,
$1^{3}A_{g}^{+}$, $1^{3}B_{1g}^{-}$, and $1^{3}A_{g}^{-}$. Of these,
optical transitions to $1^{3}B_{1g}^{+}$ and $1^{3}A_{g}^{+}$ from
the $1^{3}B_{2u}^{+}$ state are dipole forbidden within the PPP model,
because of the particle-hole symmetry inherent in the system due to
the nearest-neighbor hopping approximation. In reality, of course,
particle-hole symmetry is approximate, and, therefore, these states
are faintly visible in the excited state absorption from the $T_{1}$.
Even though in our calculations these states are absent in the computed
spectra, nevertheless, it is possible to identify them by examining
their many-particle wave functions.

Our calculated values of the excitation energies of the dipole forbidden
state $1^{3}B_{1g}^{+}$ obtained using both sets of Coulomb parameters
are in good qualitative agreement with each other, although, quantitatively
speaking, the standard parameter values are slightly larger than the
screened parameter ones. As far as the experiments are concerned,
the values of $E(1^{3}B_{1g}^{+})-E(1^{3}B_{2u}^{+}$) are available
only for naphthalene, anthracene, and tetracene, and in those cases
our calculated results appear to be slightly lower than the experimental
values. Similar trends hold for the other dipole forbidden state,
$1^{3}A_{g}^{+}$ as well, except that the standard parameter value
of $E(1^{3}A_{g}^{+})-E(1^{3}B_{2u}^{+})$ for tetracene is in excellent
agreement with the experimental ones.

\textcolor{black}{As far as the theoretical values of $E(1^{3}A_{g}^{+})-E(1^{3}B_{2u}^{+})$,
an interesting trend emerges irrespective of the parameters used,
in that with the increasing length of the oligomers up to pentacene,
this excitation energy increases. Experimentally speaking, a similar
trend is visible for all the oligomers up to tetracene, while for
the longer oligomers, the experimental values are not available. This
counterintuitive trend in the values of $E(1^{3}A_{g}^{+})-E(1^{3}B_{2u}^{+})$,
with respect to the oligomer lengths, suggests that $1^{3}A_{g}^{+}$
state is much more localized than the $1^{3}B_{2u}^{+}$ state because
of which the decrease in the value of $E(1^{3}A_{g}^{+})$, with the
increasing oligomer length, is slower as compared to $E(1^{3}B_{2u}^{+})$.
Therefore, it will be of tremendous interest to measure this value
for longer acenes as well, so that this theoretical prediction of
ours can be tested.}

For the dipole allowed state $1^{3}B_{1g}^{-}$, the excitation energies
computed by the standard parameters are larger than those computed
by screened parameters for all the oligomers. Experimental values
of these excitation energies are available for all the oligoacenes
considered here, and, on the average, much better agreement is found
with the screened parameter results.

For the $1^{3}A_{g}^{-}$ state, however, screened parameter values
of the excitation energies are larger compared to the standard ones
for naphthalene, anthracene, and tetracene (\emph{cf}. Table \ref{tab:Comparison-2-4}),
almost equal for pentacene, and smaller for hexacene and heptacene
((\emph{cf}. Table \ref{tab:Comparison-5-7}). Experimental values
of $E(1^{3}A_{g}^{-})-E(1^{3}B_{2u}^{+})$ are available up to pentacene,
and, they have disagreements amongst themselves for naphthalene and
tetracene for which multiple values have been reported. Overall, the
experimental results are in good agreement with the theory, except
for the case of pentacene, for which theoretical values are larger
than the measured one.

Next, we present a comparison of our results for individual oligomers
with the theoretical works of other authors, as well as with the experiments.

\begin{center}
\begin{table}
\protect\protect\protect\caption{Comparison of results of our triplet calculations for naphthalene,
anthracene, and tetracene, performed using the standard (std.)/screened
(scr.) parameters in PPP model, with other experimental and theoretical
results for the most important low-lying states. As explained in the
text, energies of the excited states of $^{3}B_{1g}$ and $^{3}A_{g}$
type, are given with respect to the $1^{3}B_{2u}^{+}$ state.\label{tab:Comparison-2-4}}

\begin{tabular}{ccccc}
\hline 
State  &  &  & Excitation energy (eV)  & \tabularnewline
 & Present  & work  & Experimental  & Other theoretical\tabularnewline
 & std.  & scr.  &  & \tabularnewline
\hline 
 &  &  & Naphthalene (C$_{\text{10}}$H$_{\text{8}}$)  & \tabularnewline
$1^{3}B_{2u}^{+}$  & 2.53  & 2.11  & 2.64(l)\cite{kasha1944,mcclure1949,sanche1990}  & \textcolor{black}{3.11,}\cite{thieletal2008}2.52,\cite{ramaseshaandsoos1984}2.72,\cite{hajgatoetal2009}2.18,
\cite{pariser1956}2.71,\cite{Houk}2.67,\cite{chan2007}2.98,\cite{sanche1990}3.05\cite{roos1994}\tabularnewline
$1^{3}B_{1g}^{+}$  & 1.21  & 1.16  & 1.30-1.35(l)\cite{leclercq1971}  & \textcolor{black}{1.36,}\cite{thieletal2008}1.18,\cite{hoytink1966}1.24,\cite{pariser1956}1.10,\cite{sanche1990}1.14\cite{roos1994}\tabularnewline
$1^{3}A_{g}^{+}$  & 2.06  & 1.97  & 1.97(l),\cite{leclercq1971}2.25(g)\cite{hunziker1972}  & \textcolor{black}{2.41,}\cite{thieletal2008}2.03,\cite{hoytink1966}
2.25,\cite{pariser1956}2.01,\cite{sanche1990}2.18\cite{roos1994}\tabularnewline
$1^{3}B_{1g}^{-}$  & 3.60  & 3.50  & 3.12(g),\cite{hunziker1969,wright1955}2.98(l),\cite{windsor1958}3.00(l)\cite{leclercq1971}  & 3.29,\cite{hoytink1966}3.52,\cite{pariser1956}3.24,\cite{sanche1990}2.61,\cite{roos1994}2.71\cite{hirao1996}\tabularnewline
$1^{3}A_{g}^{-}$  & 3.76  & 4.12  & 2.93(g),\cite{hunziker1972}3.63(l),\cite{windsor1958}3.10(l)\cite{leclercq1971}  & 3.35,\cite{hoytink1966} 3.63,\cite{pariser1956}2.76,\cite{sanche1990}2.81,\cite{roos1994}2.92\cite{hirao1996}\tabularnewline
\hline 
 &  &  & Anthracene (C$_{\text{14}}$H$_{10}$)  & \tabularnewline
$1^{3}B_{2u}^{+}$  & 1.73  & 1.48  & 1.82(l),\cite{kasha1944,mcclure1949}1.87(g)\cite{Schiedt201(1997)}  & 1.72,\cite{Ramasesha-acene-2002}2.0,\cite{hajgatoetal2009}1.8,\cite{quartietal2011}1.66,
\cite{pariser1956}1.81,\cite{Houk}1.99,\cite{chan2007}1.45\cite{lipari1975}\tabularnewline
$1^{3}B_{1g}^{+}$  & 1.22  & 1.10  & 1.40(l)\cite{leclercq1971}  & 1.18,\cite{pariser1956}1.35\cite{hirao1996}1.42,\cite{lipari1975}\tabularnewline
$1^{3}A_{g}^{+}$  & 2.31  & 2.11  & 2.65(l),\cite{windsor1958}2.40(l)\cite{leclercq1971}  & 2.32,\cite{pariser1956}2.62,\cite{hirao1996}3.70\cite{lipari1975} \tabularnewline
$1^{3}B_{1g}^{-}$  & 3.30  & 3.09  & 3.24(g),\cite{Allan}3.07(g),\cite{wright1955}2.92(l)\cite{leclercq1971,windsor1958}  & 3.35,\cite{lipari1975} 3.28,\cite{pariser1956}4.65,\cite{tomasello1982}2.74\cite{hirao1999}\tabularnewline
$1^{3}A_{g}^{-}$  & 3.84  & 3.96  & 3.77(l)\cite{leclercq1971}  & 4.16,\cite{lipari1975} 3.32,\cite{pariser1956}4.24,\cite{tomasello1982}3.03\cite{hirao1999}\tabularnewline
\hline 
 &  &  & Tetracene(C$_{\text{18}}$H$_{12}$)  & \tabularnewline
$1^{3}B_{2u}^{+}$  & 1.25  & 1.11  & 1.25(s)\cite{tomkiewiczandavakian1971}1.27(l),\cite{McGlynn1964}1.30(l)\cite{Sabbatini86(1982)}  & 1.22,\cite{patiandramasesha2014}1.39,\cite{hajgatoetal2009}1.12,\cite{quartietal2011}1.33,
\cite{hall1952}1.10, \cite{pariser1956}1.20,\cite{Houk}1.51\cite{chan2007}\tabularnewline
$1^{3}B_{1g}^{+}$  & 1.09  & 1.04  & 1.29(l)\cite{leclercq1971}  & 1.24,\cite{lipari1975}0.763,\cite{pariser1956}0.92,\cite{hirao1999}\tabularnewline
$1^{3}A_{g}^{+}$  & 2.55  & 2.42  & 2.58(l),\cite{leclercq1971,burdettandbardeen2012}2.60(l)\cite{windsor1958}  & 2.65,\cite{pariser1956}2.93\cite{hirao1999}\tabularnewline
$1^{3}B_{1g}^{-}$  & 3.05  & 2.79  & 2.92(g),\cite{wright1955}2.69(l),\cite{windsor1958}2.68(l),\cite{leclercq1971}2.55(l)\cite{pavlopoulos1972}  & 3.07,\cite{patiandramasesha2014}2.97,\cite{lipari1975}3.16,\cite{pariser1956}2.51\cite{hirao1999}\tabularnewline
$1^{3}A_{g}^{-}$  & 3.82  & 3.91  & 3.95(l),\cite{windsor1958}3.66(l),\cite{leclercq1971}3.01(l)\cite{pavlopoulos1972}  & 3.82,\cite{patiandramasesha2014}4.23,\cite{lipari1975} 3.39,\cite{pariser1956}3.26\cite{hirao1999}\tabularnewline
\hline 
\end{tabular}
\end{table}

\par\end{center}

\begin{center}
\begin{table}
\protect\protect\protect\caption{Comparison of results of our triplet calculations for pentacene, hexacene,
and heptacene $(n=5-7)$ performed with the standard (std.) parameters
and the screened (scr.) parameters with other experimental and theoretical
results for the most important low-lying states. The energies of the
$1^{3}B_{1g}^{-}$ and $1^{3}A_{g}^{-}$ states are given with respect
to the $1^{3}B_{2u}^{+}$ state. \label{tab:Comparison-5-7}}

\begin{tabular}{ccccc}
\hline 
State  &  &  & Excitation energy (eV)  & \tabularnewline
 & Present  & work  & Experimental  & Other theoretical\tabularnewline
 & std.  & scr.  &  & \tabularnewline
\hline 
 &  &  & Pentacene($C_{22}H_{14})$  & \tabularnewline
$1^{3}B_{2u}^{+}$  & 0.99  & 0.93  & \ \ 0.95(l),\cite{nijegorodovetal1997}0.86$\pm$0.03(s)\cite{Burgos83(1977)}\ \  & \ \ \ 0.53,\cite{lipari1975}0.91,\cite{Ramasesha-acene-2002}1.05,\cite{hajgatoetal2009}0.99,\cite{angeretal2012}0.88,\cite{zimmermanetal2010}0.99,
\cite{hall1952}0.79, \cite{pariser1956}0.78,\cite{Houk}1.16\cite{chan2007}\ \ \tabularnewline
 &  &  &  & \tabularnewline
$1^{3}B_{1g}^{+}$  & 0.96  & 0.90  &  & 0.88,\cite{pariser1956}1.06,\cite{lipari1975} 1.24\cite{zimmermanetal2010}\tabularnewline
$1^{3}A_{g}^{+}$  & 2.67  & 2.44  &  & 2.90,\cite{pariser1956} 4.15\cite{lipari1975}\tabularnewline
$1^{3}B_{1g}^{-}$  & 2.87  & 2.54  & 2.53(l),\cite{windsor1958}2.46(l)\cite{roberge1972}  & 2.76,\cite{lipari1975} 2.89\cite{pariser1956}\tabularnewline
$1^{3}A_{g}^{-}$  & 3.77  & 3.79  & 3.16(l)\cite{roberge1972}  & 4.40,\cite{lipari1975} 3.43,\cite{pariser1956}\tabularnewline
\hline 
 &  &  & Hexacene($C_{26}H_{16})$  & \tabularnewline
$1^{3}B_{2u}^{+}$  & 0.87  & 0.85  & 0.54$\pm$0.05(l)\cite{Angliker}  & 0.73,\cite{hajgatoetal2009}0.46,\cite{Houk}0.45,\cite{wudl2004}0.91\cite{chan2007}\tabularnewline
$1^{3}B_{1g}^{+}$  & 0.80  & 0.75  &  & \tabularnewline
$1^{3}A_{g}^{+}$  & 2.63  & 2.25  &  & \tabularnewline
$1^{3}B_{1g}^{-}$  & 2.70  & 2.33  & 2.25(l)\cite{Angliker}  & 2.18,\cite{patcher-dft} 2.42\cite{Angliker}\tabularnewline
$1^{3}A_{g}^{-}$  & 3.74  & 3.65  &  & \tabularnewline
\hline 
 &  &  & Heptacene($C_{30}H_{18})$  & \tabularnewline
$1^{3}B_{2u}^{+}$  & 0.73  & 0.69  &  & 0.54,\cite{hajgatoetal2009}0.24\cite{Houk}\tabularnewline
$1^{3}B_{1g}^{+}$  & 0.69  & 0.61  &  & \tabularnewline
$1^{3}A_{g}^{+}$  & 2.34  & 1.90  &  & \tabularnewline
$1^{3}B_{1g}^{-}$  & 2.58  & 2.14  & 2.14(l)\cite{neckers2008}  & \tabularnewline
$1^{3}A_{g}^{-}$  & 3.72  & 3.17  &  & \tabularnewline
\hline 
\end{tabular}
\end{table}

\par\end{center}

\subsubsection{Naphthalene}

For naphthalene and anthracene FCI calculations have been performed
which yield exact results within the chosen model Hamiltonian, and,
therefore, cannot be improved. Thus, any discrepancy in the results
with respect to the experiments has to be understood as a limitation
of the PPP model, or the parameters used to describe it.

Since we have performed calculations on isolated oligomers, therefore,
most appropriate comparison will be with the experiments performed
in the gas or liquid solution phase.

Lewis and Kasha,\cite{kasha1944} and McClure,\cite{mcclure1949}
based upon their solution based experiments, \textcolor{black}{on
phosphorescence spectra observed in EPA at 90K and in rigid glass
solution at liquid $N_{2}$ temperature, respectively, reported the
value of $E(1^{3}B_{2u}^{+}$) of naphthalene to be 2.64 eV. Exactly
the same value was also reported by Swiderek }\textcolor{black}{\emph{et
al}}\textcolor{black}{.\cite{sanche1990} based upon electron-energy
loss spectroscopy measurements of naphthalene deposited on solid argon
thin films. This value is in very good agreement with our standard
parameter based value of 2.53 eV, but is significantly larger than
the screened parameter value of 2.11 eV. As far as comparison with
other theoretical works goes, our standard parameter value of $E(1^{3}B_{2u}^{+}$)
is in almost perfect agreement with the PPP-FCI value of 2.52 eV reported
by Ramasesha and Soos,\cite{ramaseshaandsoos1984} and in good agreement
with the CCSDT(T) value of 2.72 eV reported by Hajgato }\textcolor{black}{\emph{et
al}}\textcolor{black}{.,\cite{hajgatoetal2009} first-principles DMRG
value of 2.67 eV reported by Hachmann }\textcolor{black}{\emph{et
al.}}\textcolor{black}{,\cite{chan2007} and 2.71 eV reported by Houk
}\textcolor{black}{\emph{et al}}\textcolor{black}{.,\cite{Houk}computed
using the B3LYP as exchange-correlation functional with a 6-31G{*}
basis sets. The value 3.11 eV reported by Thiel }\textcolor{black}{\emph{et
al.,}}\textcolor{black}{\cite{thieletal2008} for the excitation energy
of the $1^{3}B_{2u}^{+}$ state, computed using the coupled cluster
method CC3, with TZVP basis set, overestimates our calculated value.
On comparing with the experiments, the best estimate of the excitation
energy for the state $1^{3}B_{2u}^{+}$, from the simulations, is
2.67 eV by Hachmann }\textcolor{black}{\emph{et al.}}\textcolor{black}{\cite{chan2007}}

\textcolor{black}{For the dipole forbidden state $1^{3}B_{1g}^{+}$,
Meyer}\textcolor{black}{\emph{ et al.}}\textcolor{black}{,\cite{leclercq1971}
based upon their solution based measurement of triplet-triplet absorption
spectra of naphthalene, anthracene and tetracene using transient absorption
spectroscopy, reported $E(1^{3}B_{1g}^{+})-E(1^{3}B_{2u}^{+})$ to
be in the range 1.30-1.35 eV, a value which is in good agreement with
our standard parameter based value of 1.21 eV. Our screened parameter
value 1.16 eV underestimates the experimental value somewhat.{} Coupled-cluster
calculations, CC3 with TZVP basis set, by Thiel et al,\cite{thieletal2008}
predicts 1.36 eV as the value, which matches perfectly with the experiment.
Other reported theoretical values for this quantity is 1.14 eV calculated
using the CASPT2 approach.\cite{roos1994}}

\textcolor{black}{For the state $1^{3}A_{g}^{+}$, there are two reported
experimental values of $E(1^{3}A_{g}^{+})-E(1^{3}B_{2u}^{+})$, 1.97
eV\cite{leclercq1971} based upon the solution phase data, and 2.25
eV obtained by Hunziker\cite{hunziker1972} from the gas phase experiments,
using kinetic spectroscopy based on modulated excitation of mercury-photo-sensitized
reactions, in the visible absorption regions of the spectrum.}

\textcolor{black}{Our screened parameter value of 1.97 eV matches
exactly with the solution phase results,\cite{leclercq1971} while
the standard parameter value of 2.06 eV is in between the two experimental
values. As far as the theoretical works of other authors are concerned,
Rubio et al.\cite{roos1994} obtained 2.18 eV using the CASPT2 approach,
while the value 2.41 eV reported by Thiel }\textcolor{black}{\emph{et
al}}\textcolor{black}{.,\cite{thieletal2008} overestimates our calculated
value. On comparing with the experiments, the best estimate of the
excitation energy for the state $1^{3}A_{g}^{+}$, 1.97 eV is from
our screened parameter calculation.}

\textcolor{black}{For the long-axis polarized dipole allowed state
$1^{3}B_{1g}^{-}$, the reported experimental values of $E(1^{3}B_{1g}^{-})-E(1^{3}B_{2u}^{+})$
are in the range 2.98---3.12 eV,\cite{wright1955,windsor1958,hunziker1969,leclercq1971}
while for the short-axis polarized $1^{3}A_{g}^{-}$ state the corresponding
values fall in the range 2.93---3.63 eV.\cite{windsor1958,leclercq1971,hunziker1972}
As far as experimental techniques employed are concerned, Porter and
Wright\cite{wright1955} used flash photolysis technique to measure
the triplet absorption spectra of naphthalene, anthracene and tetracene
in the gas phase, whereas the same for naphthalene, anthracene, tetracene
and pentacene, in fluid solutions at normal temperature was measured
by Porter and Windsor\cite{windsor1958} using again the flash photolysis
and transient absorption spectroscopy. Hunziker used the absorption
of light emitted by Hg atoms in the triplet state, and the modulation
technique, to observe triplet absorption in naphthalene in the gas
phase.\cite{hunziker1969} Our calculated values for $E(1^{3}B_{1g}^{-})-E(1^{3}B_{2u}^{+})$
3.50 eV (screened parameters) and 3.60 eV (standard parameters) appear
to overestimate corresponding experimental values. Reported theoretical
values of other authors range from 2.61 eV to 3.52 eV (}\textcolor{black}{\emph{cf}}\textcolor{black}{.
Table \ref{tab:Comparison-2-4}), obtained using a variety of methods.\cite{hoytink1966,pariser1956,sanche1990,roos1994,hirao1996}
For the $1^{3}A_{g}^{-}$ state, our standard parameter value 3.76
eV for $E(1^{3}A_{g}^{-})-E(1^{3}B_{2u}^{+})$ is in excellent agreement
with 3.63 eV measured by Porter and Windsor\cite{windsor1958} in
the solution phase. Other authors have reported theoretical values
of this quantity in the range 2.76---3.63 eV.\cite{hoytink1966,pariser1956,sanche1990,roos1994,hirao1996}
As far as agreement with the gas-phase experimental values for the
excitation energies of $1^{3}B_{1g}^{-}$ and $1^{3}A_{g}^{-}$ is
concerned, the best theoretical estimates appear to be 2.71 eV and
2.92 eV , respectively, obtained by Hirao and co-workers, using MRMP
method, with cc-pVDZ basis set.\cite{hirao1996}}

\subsubsection{\textcolor{black}{Anthracene}}

\textcolor{black}{Like naphthalene, for anthracene also several experimental
and theoretical investigations have been performed over the years.
In early solution based experiments Lewis and Kasha,\cite{kasha1944}
as well as McClure\cite{mcclure1949} reported the value of $E(1^{3}B_{2u}^{+}$)
to be 1.82 eV, while in a more recent gas phase experiment using photodetachment
photoelectron spectroscopy of anthracene, Schiedt and Weinkauf\cite{Schiedt201(1997)}
measured it to be 1.87 eV. Here again the standard parameter value
of 1.78 eV is in significantly better agreement with the experiments
as compared to the screened parameter value of 1.48 eV. Similar to
the case of naphthalene, even other theoretical calculations also
match well with our standard parameter value. Ramasesha }\textcolor{black}{\emph{et
al.}}\textcolor{black}{, based upon PPP-FCI calculations reported
$E(1^{3}B_{2u}^{+}$) as 1.71 eV,\cite{ramasesha_anthracene_fci}
Houk }\textcolor{black}{\emph{et al.}}\textcolor{black}{\cite{Houk}
based upon DFT calculations, and Quarti }\textcolor{black}{\emph{et
al.\cite{quartietal2011}}}\textcolor{black}{{} relying on a variety
of quantum-chemical calculations reported it to be 1.8 eV, while Hachmann
}\textcolor{black}{\emph{et al.}}\textcolor{black}{\cite{chan2007}
and Hajgato }\textcolor{black}{\emph{et al}}\textcolor{black}{.\cite{hajgatoetal2009}
computed it as 2.0 eV. On comparing with the experiments, the best
estimate of the excitation energy for the state $1^{3}B_{2u}^{+}$,
is 1.81 eV, obtained by Houk and co-workers, using B3LYP as exchange-correlation
functional with a 6-31G{*} basis sets.\cite{Houk}}

\textcolor{black}{For the dipole forbidden state $1^{3}B_{1g}^{+}$,
the reported experimental value of $E(1^{3}B_{1g}^{+})-E(1^{3}B_{2u}^{+})$
1.40 eV\cite{leclercq1971} is closer to our standard parameter value
1.22 eV as compared to the screened parameter one. Theoretical works
of other authors lie in the range 1.18---1.42 eV.\cite{pariser1956,lipari1975,hirao1996}
For the higher dipole forbidden state $1^{3}A_{g}^{+}$, two experimental
values 2.40 eV\cite{leclercq1971} and 2.65 eV\cite{windsor1958}
based upon the liquid phase data have been reported, and both are
in better agreement with our standard parameter value 2.31 eV, as
compared to the screened one. Theoretical values reported by other
authors for $E(1^{3}A_{g}^{+})-E(1^{3}B_{2u}^{+})$ are 2.32 eV,\cite{pariser1956}
2.62 eV,\cite{hirao1996}, in good agreement with our standard parameter
results.{} On comparing with the experiments, the best estimates
of the excitation energies for the states $1^{3}B_{1g}^{+}$ and $1^{3}A_{g}^{+}$,
1.35 eV and 2.62 eV ,respectively, were obtained by Hashimoto }\textcolor{black}{\emph{et
al.}}\textcolor{black}{, using MRMP method, cc-pVDZ basis set.\cite{hirao1996}}

\textcolor{black}{For the dipole allowed state $1^{3}B_{1g}^{-}$,
the value of $E(1^{3}B_{1g}^{-})-E(1^{3}B_{2u}^{+})$ from solution
based experiments was reported to be 2.92 eV,\cite{leclercq1971,windsor1958}
while gas phase experiments report two distinct values 3.07 eV,\cite{wright1955}
and 3.24 eV.\cite{Allan}For the gas-phase experiment reporting the
larger value,\cite{Allan} the technique employed was electron energy
loss spectroscopy. Interestingly, both our screened and standard parameter
values are in good agreement with the two gas phase results.}

\textcolor{black}{For the short-axis polarized dipole allowed state
$1^{3}A_{g}^{-}$, only available experimental value of $E(1^{3}A_{g}^{-})-E(1^{3}B_{2u}^{+})$
is 3.77 eV, measured in the solution phase.\cite{leclercq1971} Our
standard parameter value of 3.84 eV is in excellent agreement with
this value, while the screened parameter value 3.96 eV overestimates
it a bit. Other theoretical values of this quantity are distributed
in the wide range 3.03---4.24 eV.\cite{lipari1975,pariser1956,tomasello1982,hirao1999}
Upon comparing with the experiments, the best estimates of the excitation
energies of the dipole-allowed states, $1^{3}B_{1g}^{-}$, and $1^{3}A_{g}^{-}$,
are 3.30 eV, and 3.84 eV , respectively, obtained in our standard
parameter calculation.}

\subsubsection{\textcolor{black}{Tetracene}}

\textcolor{black}{For tetracene, the value of $E(1^{3}B_{2u}^{+}$)
1.25 eV was obtained by Tomkiewicz }\textcolor{black}{\emph{et al.\cite{tomkiewiczandavakian1971}}}\textcolor{black}{{}
by measuring the singlet-triplet spectrum in solid phase as observed
in the delayed fluorescence from a tetracene crystal at room temperature.
Liquid phase experiments of Mc Glynn }\textcolor{black}{\emph{et al.
}}\textcolor{black}{\cite{McGlynn1964} yielded the value 1.27 eV,
while Sabbatini }\textcolor{black}{\emph{et al.}}\textcolor{black}{\cite{Sabbatini86(1982)}
measured it to be 1.30 eV, from the emission spectra of the triplet
states of naphthalene, anthracene and tetracene in acetonitrile solution.
These experimental values are in closer agreement with our standard
parameter based value of 1.25 eV, as compared to the screened parameter
value 1.11 eV. In a recent remarkable theoretical work, Pati and Ramasesha
managed to perform PPP-FCI calculations for anthracene,\cite{patiandramasesha2014}
and their $E(1^{3}B_{2u}^{+}$) value of 1.22 eV is in excellent agreement
with our standard parameter result obtained using the QCI approach.
Theoretical calculations by other authors predict this energy to be
in the range 1.10---1.51 eV.\cite{hajgatoetal2009,chan2007,Houk,pariser1956,hall1952,quartietal2011}
In terms of the level of electron correlations taken into account,
Pati and Ramasesha's\cite{patiandramasesha2014} PPP-FCI calculations{}
are the best, although our PPP-QCI value obtained with standard parameters
is in marginally better agreement with the experiments. }

\textcolor{black}{For the dipole forbidden state $1^{3}B_{1g}^{+}$,
only one reported experimental value of $E(1^{3}B_{1g}^{+})-E(1^{3}B_{2u}^{+})$
1.29 eV\cite{leclercq1971} is based upon liquid phase measurement.
Our calculated values are lower than this, with the standard parameter
value of 1.09 eV, being somewhat better than the screened parameter
value 1.04 eV. Among the theoretical works of other authors, only
the value 1.24 eV computed by Lipari and Duke\cite{lipari1975} is
in better agreement with the experiments as compared to our results,
and appears to be the best theoretical estimate. For the other dipole
forbidden state $1^{3}A_{g}^{+}$, solution based measurements yield
values of $E(1^{3}A_{g}^{+})-E(1^{3}B_{2u}^{+})$ as 2.58 eV,\cite{leclercq1971}
and 2.60 eV,\cite{windsor1958} while Burdett and Bardeen,\cite{burdettandbardeen2012}
using transient absorption and photoluminescence experiments on crystalline
tetracene obtained it to be 2.58 eV. Thus, experimental values are
close to each other, and also are in close agreement with our standard
parameter value 2.55 eV, which also happens to be in best agreement,
among all theoretical values, with the experiments. Other reported
theoretical values of this quantity overestimate the experiments (}\textcolor{black}{\emph{cf.}}\textcolor{black}{{}
Table \ref{tab:Comparison-2-4}).}

\textcolor{black}{As in the case of anthracene, for the first dipole
allowed state $1^{3}B_{1g}^{-}$, significant variation in the experimental
values of $E(1^{3}B_{1g}^{-})-E(1^{3}B_{2u}^{+})$ is observed. Porter
and Wright \cite{wright1955} measured it to be 2.92 eV in gas phase,
which is in excellent agreement with our standard parameter value
of 3.05 eV. Pavlopoulos\cite{pavlopoulos1972} reported experimental
value of 2.55 eV based upon solution phase measurements of the triplet-triplet
absorption spectrum of tetracene using continuous wave (cw) laser
excitation, while values obtained in other liquid phase experiments
are 2.68 eV,\cite{leclercq1971} and 2.69 eV.\cite{windsor1958} These
values, quite expectedly, are in better agreement with our screened
parameter value 2.79 eV, as compared to the standard parameter result.
As far as theoretical results of other authors are concerned, recent
PPP-FCI value 3.07 eV reported by Pati and Ramasesha\cite{patiandramasesha2014}
is in excellent agreement with our standard parameter value, while
other reported values in the literature fall in the range 2.51---3.16
eV.\cite{pariser1956,lipari1975,hirao1999}}

\textcolor{black}{For the higher dipole allowed state $1^{3}A_{g}^{-}$,
even though all the experimental measurements of $E(1^{3}A_{g}^{-})-E(1^{3}B_{2u}^{+})$
were performed in the liquid phase, yet they exhibit significant variation
with reported values 3.01 eV,\cite{pavlopoulos1972} 3.66 eV,\cite{leclercq1971}
and 3.95 eV.\cite{windsor1958} Because of this variation, both our
standard and screened parameter values are in good agreement with
one experimental value or another (}\textcolor{black}{\emph{cf}}\textcolor{black}{.
Table \ref{tab:Comparison-2-4}). As shown in Table \ref{tab:Comparison-2-4},
theoretical values of this quantity computed by other authors also
exhibit wide variation, with our standard parameter results being
in perfect agreement with the PPP-FCI values reported by Pati and
Ramasesha.\cite{patiandramasesha2014}On comparing with the liquid
phase experiments, the best theoretical estimates of the excitation
energies for the states $1^{3}B_{1g}^{-}$ and $1^{3}A_{g}^{-}$,
2.79 eV and 3.91eV, respectively, are from our screened parameter
calculations.}

\subsubsection{\textcolor{black}{Pentacene}}

\textcolor{black}{Pentacene is one of the most widely studied polyacene,
and extremely important from the point of view of singlet fission
problem, currently being studied vigorously.\cite{singlet-fission-michl,singlet-fission-michl2}
Nijegorodov }\textcolor{black}{\emph{et al}}\textcolor{black}{.\cite{nijegorodovetal1997}
measured the value of $E(1^{3}B_{2u}^{+}$) to be 0.95 eV in a liquid
phase experiment using UV-visible absorption spectroscopy. Burgos
}\textcolor{black}{\emph{et al. }}\textcolor{black}{\cite{Burgos83(1977)},
on the other hand, obtained the value 0.86 eV{} by performing fluorescence
measurements of pentacene in a solid phase experiment. Both our standard
parameter value 0.99 eV, and screened parameter value of 0.93 eV,
are in good agreement with the liquid phase measurement. Furthermore,
quite expectedly, the screened parameter value is closer to the solid
state measurement as compared to the standard parameter result. A
large number of theoretical calculations have been performed for this
quantity for pentacene, and they range from a low value of 0.53 eV
obtained in a CNDO/S2-CI calculation,\cite{lipari1975}to a high value
of 1.16 eV obtained from a DMRG calculation.\cite{chan2007} However,
several other calculations predict values in good agreement with our
work, and with the experiments.\cite{Ramasesha-acene-2002,angeretal2012,zimmermanetal2010,hall1952}On
comparing with the liquid phase experiments, besides our own screened
parameter results, the best theoretical values of the excitation energy
of the state $1^{3}B_{2u`}^{+}$ were predicted by PPP-DMRG calculations
of Raghu }\textcolor{black}{\emph{et al}}\textcolor{black}{.,\cite{Ramasesha-acene-2002}
and time-dependent density-functional-theory (TDDFT) calculations
of Anger }\textcolor{black}{\emph{et al}}\textcolor{black}{.\cite{angeretal2012}
On the other hand TDDFT results of Zimmerman }\textcolor{black}{\emph{et
al}}\textcolor{black}{.,\cite{zimmermanetal2010} agree well with
solid phase experimental value.}

\textcolor{black}{To the best of our knowledge, no experimental measurements
of the dipole forbidden states $1^{3}B_{1g}^{+}$ and $1^{3}A_{g}^{+}$
exist for pentacene, therefore, we will restrict the comparison of
our results on these states to the theoretical works of other authors.
Pariser\cite{pariser1956} computed the value of $E(1^{3}B_{1g}^{+})-E(1^{3}B_{2u}^{+})$
to be 0.88 eV, while Lipari and Duke predicted it to be 1.06 eV, both
of which are in reasonable agreement with our computed values. The
TDDFT value of 1.24 eV calculated by Zimmerman }\textcolor{black}{\emph{et
al}}\textcolor{black}{.\cite{zimmermanetal2010} for the quantity
is substantially larger than our prediction. For the $1^{3}A_{g}^{+}$
state our calculated values of $E(1^{3}A_{g}^{+})-E(1^{3}B_{2u}^{+})$
2.44 eV (screened), and 2.67 eV (standard), are lower compared to
the PPP-SCI value 2.90 eV reported by Pariser,\cite{pariser1956}
but significantly smaller compared to 4.15 eV calculated by Lipari
and Duke.\cite{lipari1975} Given the uncertainty in the theoretical
excitation energies of these dipole forbidden states, it will be good
if measurements are performed on these states of pentacene in future.}

\textcolor{black}{For the long-axis polarized dipole-allowed state
$1^{3}B_{1g}^{-}$, two liquid phase experimental values of $E(1^{3}B_{1g}^{-})-E(1^{3}B_{2u}^{+})$
2.46 eV,\cite{roberge1972} obtained using flash photolysis, absorption
spectrum of triplet pentacene in benzene and cyclohexane at $50\textdegree C$,
and 2.53 eV reported by Porter and Windsor,\cite{windsor1958} are
in excellent agreement with our calculated screened parameter value
2.54 eV, while the standard parameter approach predicts a larger value.
Theoretical calculations by other authors\cite{pariser1956,lipari1975}
(}\textcolor{black}{\emph{cf}}\textcolor{black}{. Table \ref{tab:Comparison-5-7})
predict values larger than the experimental one, and are in better
agreement with our standard parameter calculation. Therefore, as far
as the agreement with the experimental results is concerned, the best
estimate of the excitation energy of the state $1^{3}B_{1g}^{-}$,
is 2.54 eV predicted by our screened parameter calculation. }

\textcolor{black}{For the other dipole allowed state $1^{3}A_{g}^{-}$,
we have an anomaly vis-a-vis experiments in that the only available
experimental value of $E(1^{3}A_{g}^{-})-E(1^{3}B_{2u}^{+})$ 3.16
eV,\cite{roberge1972} is substantial smaller than all the calculated
theoretical values, including our own. Therefore, it will be quite
useful if an experiment could be repeated on pentacene to ascertain
the excitation energy of this state.}

\subsubsection{\textcolor{black}{Hexacene and Heptacene}}

\textcolor{black}{Very few experiments have been performed on the
triplet states of higher acenes such as hexacene and heptacene, perhaps
because of difficulties associated with their synthesis. As a result,
the number of theoretical calculations on these compounds are relatively
small. For hexacene, based on a liquid phase experiment, Angliker
}\textcolor{black}{\emph{et al}}\textcolor{black}{.\cite{Angliker},
using the absorption spectra of triplet hexacene, by flash photolysis,
estimated the value of $E(1^{3}B_{2u}^{+}$) to be 0.54 eV, which
is substantially smaller than our computed values 0.85 eV (screened)
and 0.87 eV (standard). This reported experimental value is also not
in agreement with the theoretical works of other authors whose values
are either significantly smaller than it, or larger than it (}\textcolor{black}{\emph{cf}}\textcolor{black}{.
Table \ref{tab:Comparison-5-7}). In the same experiment, Angliker
}\textcolor{black}{\emph{et al}}\textcolor{black}{.\cite{Angliker}}\textcolor{black}{\emph{
}}\textcolor{black}{measured $E(1^{3}B_{1g}^{-})-E(1^{3}B_{2u}^{+})$
to be 2.25 eV which is actually in very good agreement with our screened
parameter value 2.33 eV. Angliker }\textcolor{black}{\emph{et al}}\textcolor{black}{.\cite{Angliker}
also calculated the value of this excitation energy using the PPP-SCI
approach using a different set of Coulomb parameters, and obtained
the value 2.42 eV, which is in good agreement with our screened parameter
result. DFT based calculations of Nguyen }\textcolor{black}{\emph{et
al}}\textcolor{black}{.\cite{patcher-dft} predict the value of this
energy as 2.18 eV, which is close to the experimental value, but somewhat
smaller than our screened parameter result. Thus, on comparing with
the experiments, the best estimate of the excitation energy for the
state $1^{3}B_{2u}^{+}$, 0.45-0.46 eV by Houk and co-workers,\cite{Houk,wudl2004}
using B3LYP method, 6-31G{*} basis set, and for the state $1^{3}B_{1g}^{-}$,
2.33 eV obtained in our screened parameter calculation. }

\textcolor{black}{For heptacene, to the best of our knowledge, no
experimental measurements of $E(1^{3}B_{2u}^{+}$) exist. Theoretical
calculations by Hajgato }\textcolor{black}{\emph{et a}}\textcolor{black}{l.\cite{hajgatoetal2009}
predict it to be 0.54 eV, while the DFT based work of Houk }\textcolor{black}{\emph{et
al}}\textcolor{black}{.\cite{Houk} estimates a rather small value
of 0.24 eV. From Table \ref{tab:Comparison-5-7} it is obvious that
our calculated values are larger than these results, with a somewhat
better agreement between the work of Hajgato }\textcolor{black}{\emph{et
a}}\textcolor{black}{l.\cite{hajgatoetal2009} and our screened parameter
estimate. However, liquid phase experiment on the dipole allowed $1^{3}B_{1g}^{-}$
state of heptacene performed by Mondal }\textcolor{black}{\emph{et
al}}\textcolor{black}{.\cite{neckers2008} using nanosecond laser
flash photolysis and femtosecond pump-probe UV-vis spectrometry, measured
$E(1^{3}B_{1g}^{-})-E(1^{3}B_{2u}^{+})$ to be 2.14 eV, which is in
exact agreement with our screened parameter result, thus giving us
confidence about the quality of our calculations even for a relatively
longer oligoacene such as heptacene.}

\subsection{Triplet Absorption Spectrum{\normalsize{}{}\label{sub:Triplet-excited-state}}}

\textcolor{black}{First singlet excited state $S_{1}$ ($1^{1}B_{2u}^{+}$,
in the present case) of conjugated molecules frequently decays to
the first triplet excited state $T_{1}$ ($1^{3}B_{2u}^{+}$, in the
present case) located below $S_{1}$, through non-radiative inter-system
crossing (ISC}\textcolor{black}{\emph{), }}\textcolor{black}{as shown
in Fig. \ref{fig:PA}. In case of single crystals or thin films, this
transition is believed to occur via singlet fission,\cite{thorsmolleetal2009}
whose mechanism is also an area of intense research.\cite{singlet-fission-michl,singlet-fission-michl2}
Once the triplet state is attained, usual optical absorption experiments
can be conducted to probe higher triplet states. In case of oligoacenes,
according to the electric-dipole selection rules of the $D_{2h}$
point group, the transitions from the $1^{3}B_{2u}^{+}$ to $^{3}B_{1g}^{-}$
type of states is caused by the long-axis ($x$-axis) polarized photons,
whereas the transitions to the $^{3}A_{g}^{-}$ type states by the
short-axis ($y$-axis) polarized ones (}\textcolor{black}{\emph{cf}}\textcolor{black}{.
Fig. \ref{fig:PA}). In figs. \ref{fig:acene2_pa}---\ref{fig:acene7_pa},
we present the triplet absorption spectra of naphthalene to heptacene,
while the wave functions of the excited states contributing to peaks
in the spectra, along with corresponding transition dipole moments,
are presented in Tables \ref{tab:acene2-pa-std}---\ref{tab:acene7-pa-scr}
of Appendix \ref{sec:appendixB}. Next, we summarize the salient features
of the triplet absorption spectra, and briefly discuss the many-particle
wave functions of the excited states contributing to it. }

\subsubsection{\textcolor{black}{Dipole forbidden spectrum}}

\textcolor{black}{As mentioned earlier, $x$-polarized $1^{3}B_{1g}^{+}$
state, and $y$-polarized $1^{3}A{}_{g}^{+}$ state do not contribute
to the computed spectra because they are dipole forbidden due to the
particle-hole symmetry. Nevertheless, in experimental spectra they
contribute faint peaks, therefore, we are presenting a discussion
of their many-particle wave functions. We note that the wave functions
of both these states are dominated by singly excited configurations.
On adopting the notation that $H$ denotes the highest occupied molecular
orbital (HOMO), and $L$ denotes the lowest unoccupied molecular orbital
(LUMO),{} we find that the wave function of the $1^{3}B_{1g}^{+}$
state has dominant contributions from $\vert H\rightarrow L+2\left\rangle \right.\:\mbox{and }c.c.$
($c.c.$ denotes the electron-hole conjugate configuration), for naphthalene
and anthracene, and from single excitations $\vert H\rightarrow L+1\rangle\;\mbox{and }c.c.$,
for tetracene up to heptacene, irrespective of the choice of Coulomb
parameters. On the other hand, $1^{3}A_{g}^{+}$ state has dominant
contributions from excitations $\vert H\rightarrow L+3\left\rangle \right.+c.c.$
for the case of naphthalene and anthracene, from the single excitations
$\vert H\rightarrow L+4\rangle+c.c.$ for tetracene up to hexacene,
but from double excitation $\vert H\rightarrow L;H-1\rightarrow L+1\rangle$,
for heptacene, irrespective of the choice of Coulomb parameters. Dominant
contribution from double excitation also points to the increased electron-correlation
effects contributing to this state. }

\subsubsection{\textcolor{black}{Dipole allowed spectrum}}

\textcolor{black}{In a recent work, we calculated the triplet absorption
spectra of longer oligoacenes namely octacene, nonacene and decacene,
with an aim to understand their experimentally measured optical absorption.\cite{chakrabortyandshukla2013}
In this section we analyze the similarities and differences between
the triplet absorption spectra of shorter acenes (naphthalene to heptacene),
when compared to those of the longer ones (octacene, nonacene and
decacene). }

\textcolor{black}{As far as the similarities between the short and
the long acenes are concerned, irrespective of the length of acene,
triplet spectrum of a given acene computed using the screened parameters
exhibits a red shift as compared to the one calculated using the standard
parameters. Moreover, the $x$-polarized (long-axis polarized) spectra,
corresponding to the absorption into the $^{3}B_{1g}^{-}$ type of
states is quite intense, while the $y-$polarized absorption into
the $^{3}A_{g}^{-}$ type states is found to be rather weak. In addition,
it is evident from the triplet absorption spectra of each oligoacene
calculated using either of the Coulomb parameters, that there are
two prominent intense $x$-polarized peaks which are well separated
in energy ($>$1.5 eV). Furthermore, the many-particle wave function
of the Peak I, corresponding to the state, $1^{3}B_{1g}^{-}$ is dominated
by singly-excited configurations $\vert H\rightarrow L+1\left\rangle \right.+c.c.$,
for the oligomers from tetracene up to decacene, irrespective of the
choice of Coulomb parameters, except for the case of naphthalene and
anthracene, where singly-excited configurations $\vert H\rightarrow L+2\left\rangle \right.+c.c.$
dominate the many-particle wave function of the same peak. Other peaks
in the spectra correspond to either $x$ or $y$ polarized transitions
to the higher excited states of the system, information about which
is presented in the tables in Appendix \ref{sec:appendixB}.}

\textcolor{black}{Next, we elucidate some of the important differences
we have observed between the computed triplet spectra of short and
long acenes. In our work on the longer acenes we found that, out of
the two intense $x$-polarized peaks, the first one (peak I) is the
most intense one in the screened parameter calculations, while the
second one (peak IV or V) has maximum intensity in the standard parameter
results.\cite{chakrabortyandshukla2013} However, in the present work
on shorter acenes, the trend is somewhat different in that, irrespective
of the choice of the Coulomb parameters employed, of the two strong
peaks, first peak (I) is always the second most intense one, while,
in all the cases, the most intense peak is the second one (III, IV
or V, depending on the size of the oligomer). As far as the excited
state wave functions are concerned, for a longer acene of length $n$
(=8--10), in the screened parameter calculations, the wave function
of the state corresponding to the second intense peak (IV or V) consists
of $\vert H\rightarrow L;\, H-(n/2-1)\rightarrow L\left\rangle \right.+c.c.$
for $n\equiv\mbox{even}$, and $\vert H\rightarrow L;\, H-(n-1)/2\rightarrow L\left\rangle \right.+c.c.$,
for $n\equiv\mbox{odd}$.\cite{chakrabortyandshukla2013} With the
standard parameters, for the same oligomer the main configurations
contributing to the peak were $\vert H\rightarrow L;\, H-n/2\rightarrow L\left\rangle \right.+c.c.$
for $n\equiv\mbox{even}$, and $\vert H\rightarrow L;\, H-(n-1)/2\rightarrow L\left\rangle \right.+c.c.$,
for $n\equiv\mbox{odd}$.\cite{chakrabortyandshukla2013} Whereas,
in the current work on shorter acenes ($n=2$--7), irrespective of
the choice of Coulomb parameters, the wave function corresponding
to the second intense peak (III, IV or V) mainly consists of double
excitations $\vert H\rightarrow L;\, H-n/2\rightarrow L\left\rangle \right.+c.c.$
for $n\equiv\mbox{even}$, and $\vert H\rightarrow L;\, H-(n-1)/2\rightarrow L\left\rangle \right.+c.c.$,
for $n\equiv\mbox{odd}$. }

\begin{center}
\begin{figure}[H]
\protect\protect\protect\caption{\textcolor{black}{Triplet optical absorption spectra of naphthalene
from the $1^{3}B_{2u}^{+}$ state computed using the standard parameters
(panel (a)), and the screened parameters (panel (b)), and a uniform
line width of 0.1 eV. The polarization directions ($x$ or $y$) are
indicated by the subscripts attached to the peak labels. \label{fig:acene2_pa}}}

\centering{}\includegraphics[clip,width=8cm]{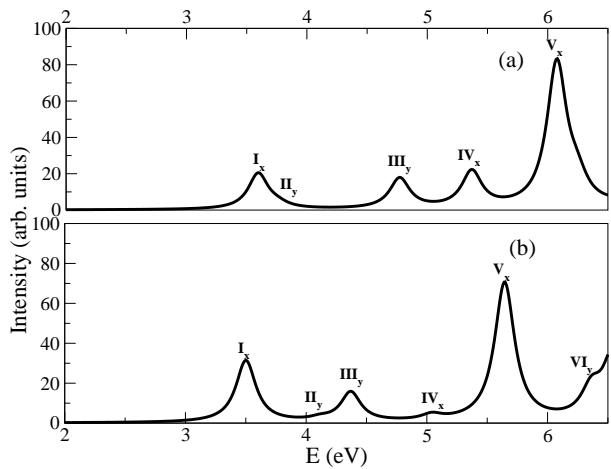} 
\end{figure}

\par\end{center}

\begin{center}
\begin{figure}[H]
\protect\protect\protect\caption{\textcolor{black}{Triplet optical absorption spectra of anthracene
from the $1^{3}B_{2u}^{+}$ state computed using the standard parameters
(panel (a)), and the screened parameters (panel (b)). The rest of
the information is same as in the caption of Fig. \ref{fig:acene2_pa}.\label{fig:acene3_pa}}}

\centering{}\includegraphics[clip,width=8cm]{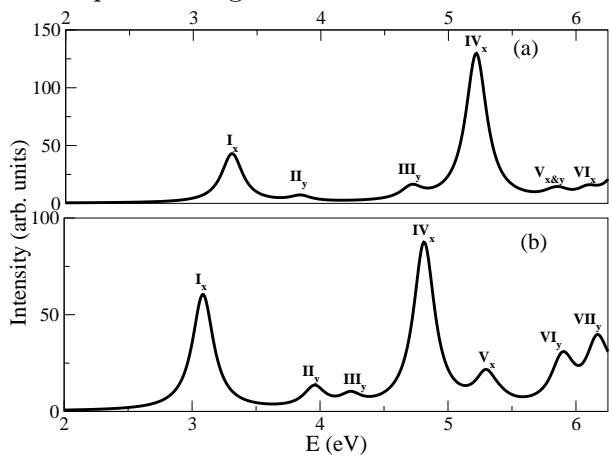} 
\end{figure}

\par\end{center}

\begin{center}
\begin{figure}[H]
\protect\protect\protect\caption{\textcolor{black}{Triplet optical absorption spectra of tetracene
from the $1^{3}B_{2u}^{+}$ state computed using the standard parameters
(panel (a)), and the screened parameters (panel (b)). The rest of
the information is same as in the caption of Fig. \ref{fig:acene2_pa}.\label{fig:acene4_pa}}}

\centering{}\includegraphics[clip,width=8cm]{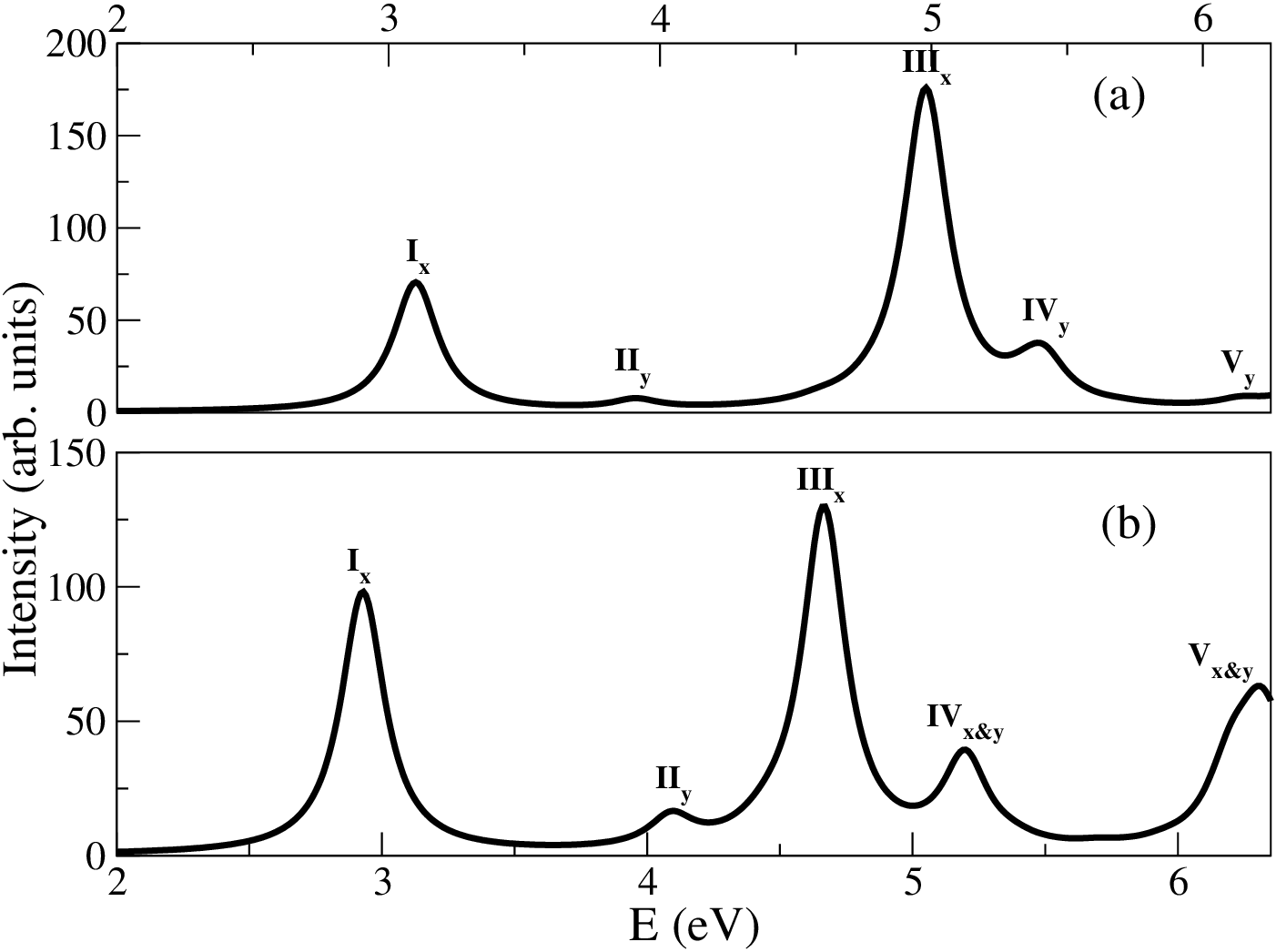} 
\end{figure}

\par\end{center}

\begin{center}
\begin{figure}[H]
\protect\protect\protect\caption{\textcolor{black}{Triplet optical absorption spectra of pentacene
from the $1^{3}B_{2u}^{+}$ state computed using the standard parameters
(panel (a)), and the screened parameters (panel (b)). The rest of
the information is same as in the caption of Fig. \ref{fig:acene2_pa}.
\label{fig:acene5_pa}}}

\centering{}\includegraphics[clip,width=8cm]{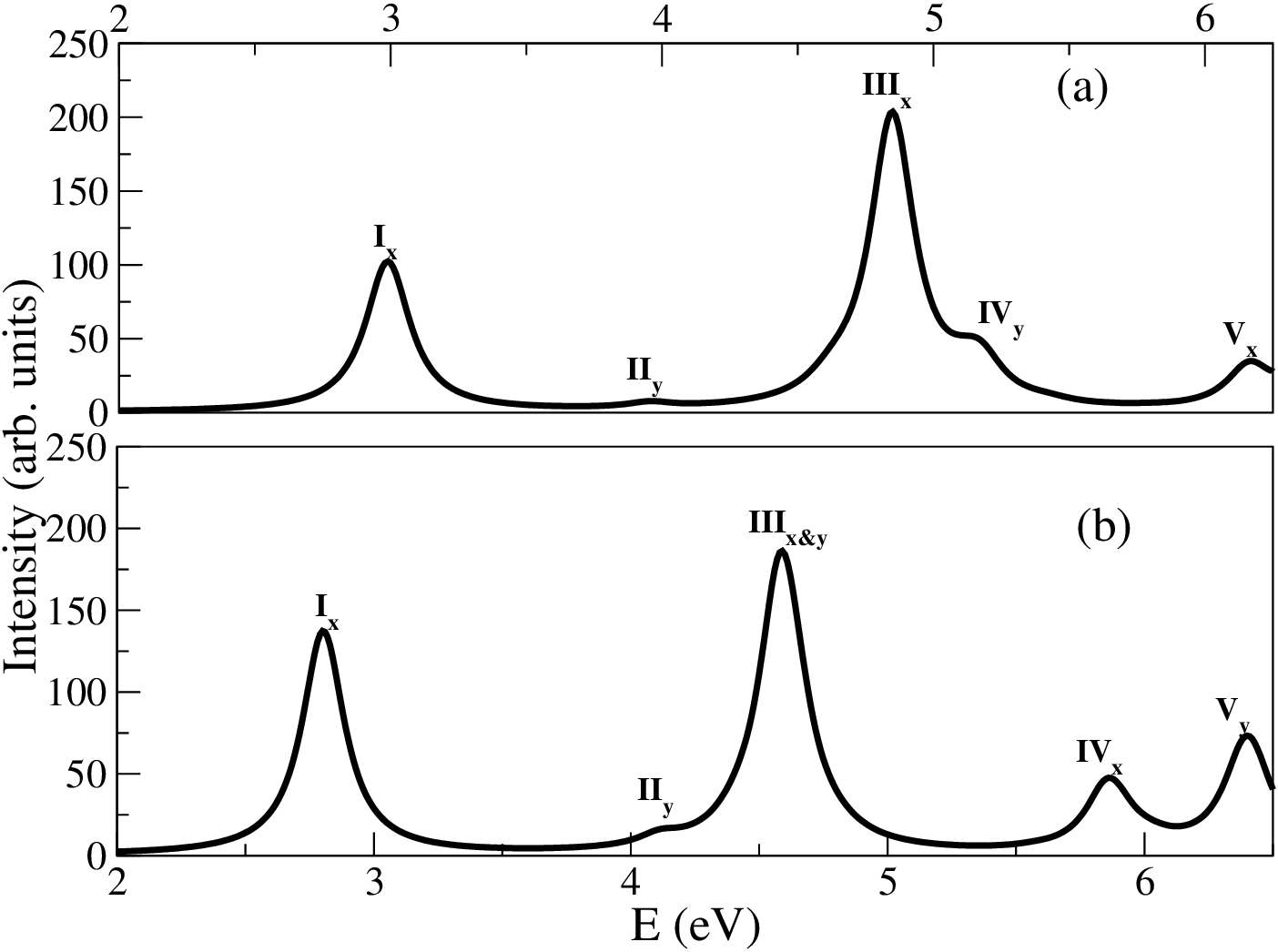} 
\end{figure}

\par\end{center}

\begin{center}
\begin{figure}[H]
\protect\protect\protect\caption{\textcolor{black}{Triplet optical absorption spectra of hexacene from
the $1^{3}B_{2u}^{+}$ state computed using the standard parameters
(panel (a)), and the screened parameters (panel (b)). The rest of
the information is same as in the caption of Fig. \ref{fig:acene2_pa}.\label{fig:acene6_pa}}}

\centering{}\includegraphics[clip,width=8cm]{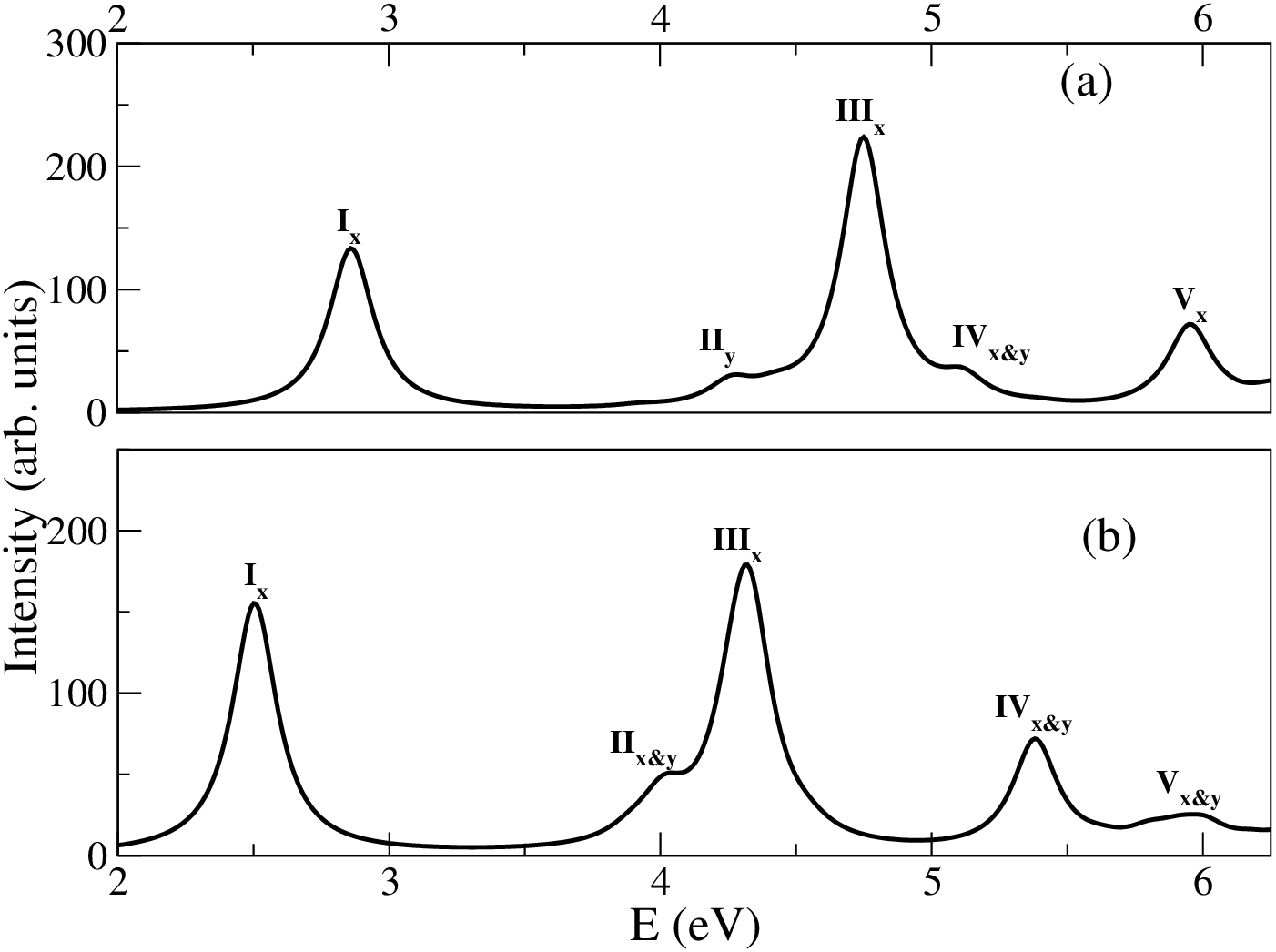} 
\end{figure}

\par\end{center}

\begin{center}
\begin{figure}[H]
\protect\protect\protect\caption{\textcolor{black}{Triplet optical absorption spectra of heptacene
from the $1^{3}B_{2u}^{+}$ state computed using the standard parameters
(panel (a)), and the screened parameters (panel (b)). The rest of
the information is same as in the caption of Fig. \ref{fig:acene2_pa}.\label{fig:acene7_pa}}}

\centering{}\includegraphics[width=8cm]{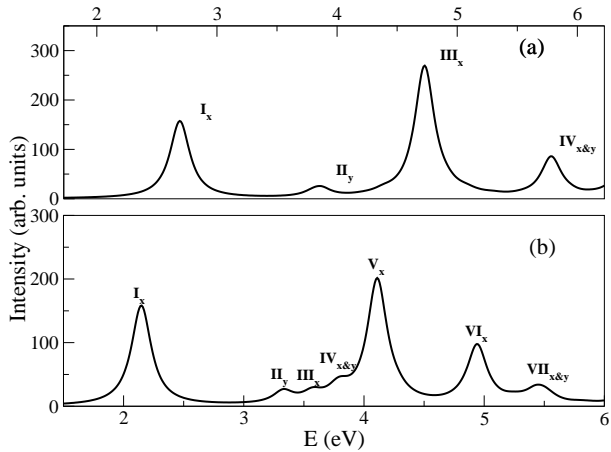} 
\end{figure}

\par\end{center}

\subsection{Comparison with poly-(para)phenylene vinylene chains}

Several years back, one of us performed a systematic study of triplet
states and triplet optical absorption in oligomers of poly-(para)phenylene
vinylene (PPV).\cite{Shukla-PPV} Next we compare our earlier results
on PPV with the present ones on oligoacenes. In finite PPV chains
excitation energy of $T_{1}$ ($1^{3}B_{u}^{+}$ state) exhibited
almost complete saturation starting from the two-unit oligomer, implying
that $T_{1}$ is highly localizd in PPV, and essentially has a Frenkel
exciton character.\cite{Shukla-PPV} However, it is obvious from Tables
\ref{tab:Comparison-2-4} and \ref{tab:Comparison-5-7}, and also
from our earlier work on longer acenes,\cite{chakrabortyandshukla2013}
that the excitation energy of $T_{1}$ ($1^{3}B_{2u}^{+}$) for oligoacenes
exhibits no signs of saturation with increasing length, implying that,
if $T_{1}$ for polyacene is an exciton in the thermodynamic limit,
it is a highly delocalized one, similar to a Wannier exciton.

When we compare the triplet absorption spectra of PPV chains\cite{Shukla-PPV}
with those of oligoacenes, following differences emerge: (a) the first
peak in the absorption spectrum of PPV chains is always the most intense
one, while in case of oligoacenes the first peak is always the second
most intense peak, and (b) successive peaks in case of PPV chains
exhibit a pattern of alternating high and low intensities, while no
such pattern is visible in case of oligoacenes.

In an earlier work, Shimoi and Mazumdar\cite{sumit-shimoi} had argued
that the $m^{3}A_{g}$ state of conjugated polymers, which is the
first excited state with a strong dipole coupling to $T_{1}$, is
also an exciton, and, therefore, the quantity $E(m^{3}A_{g})-E(1^{1}B_{u})$
can serve as a lower-limit estimate of the binding energy of the optical
exciton ($1^{1}B_{u}$). For the case of oligoacenes the first optical
state is $1^{1}B_{2u}^{+}$, while $m^{3}A_{g}$ is the $1^{3}B_{1g}^{-}$
state. Therefore, using the values of $E(1^{3}B_{1g}^{-})$ computed
in this work, and $E(1^{1}B_{2u})$ calculated in our earlier work,\cite{sony-acene-lo}
for heptacene we obtain the values of $E(m^{3}A_{g})-E(1^{1}B_{u})$
to be 0.68 eV (standard) and 0.59 eV (screened). Our screened parameter
value of 0.59 eV obtained for polyacenes is lower than 0.68 eV which
we estimated for PPV using the same Coulomb parameters in the PPP
model.\cite{Shukla-PPV} Furthermore, if we examine the behavior of
$E(m^{3}A_{g})-E(1^{1}B_{u})$ as a function of the length of oligoacene,
with both sets of Coulomb parameters a decreasing trend is observed.
Therefore, it will be of considerable interest to know whether, in
the thermodynamic limit, exciton binding energy defined in this manner
saturates to a finite value, or simply vanishes.

\section{Conclusion}

\label{sec:conclusion}

In this work we presented large-scale electron-correlated calculations
of the triplet states, as well as triplet-triplet absorption spectra,
of oligoacenes ranging from naphthalene to heptacene. Generally very
good agreement was observed between our calculations, and available
experimental results. For the case of hexacene and heptacene, not
much experimental data is available, therefore, our results on those
oligomers could be tested in future experiments. As far as triplet
absorption spectrum is concerned, our most important prediction is
the presence of two long-axis polarized well-separated peaks, which
we hope will be tested in future experiments on oriented samples of
oligoacenes. We have also presented a detailed analysis of the wave
functions of the important triplet excited states.

When compared to $T_{1}$ in PPV, our calculations find that the $T_{1}$
in oligoacenes is significantly delocalized, and has a Wannier exciton
like character. Furthermore, we have observed that the triplet absorption
spectra of polyacenes are qualitatively different as compared to those
of the PPV chains. We have also presented numerical estimates of the
optical exciton binding energy in oligoacenes, and found it to be
less than that in PPV chains. However, these estimates can be significantly
improved by performing calculations on longer oligoacenes. Moreover,
it will also be of interest to probe the nonlinear optical properties
of polyacenes, a topic which we intend to pursue in future works in
our group. 
\begin{acknowledgments}
One of us (H.C.) thanks Council of Scientific and Industrial Research
(CSIR), India for providing a Senior Research Fellowship (SRF). We
thankfully acknowledge the computational resources (PARAM-YUVA) provided
for this work by Center for Development of Advanced Computing (C-DAC),
Pune.

\newpage{} 
\end{acknowledgments}

\appendix

\section{\textcolor{black}{Convergence of Lowest Triplet Excitation Energies}}

\textcolor{black}{\label{sec:appendixA}}

\subsection{\textcolor{black}{Influence of PPP model parameters on excitation
energies}}

\textcolor{black}{\label{appa-parameters}}

\textcolor{black}{In Fig. \ref{fig:Convergence-model-parameters}
we present the plots of the excitation energy of the lowest triplet
state, $1^{3}B_{2u}^{+}$, of naphthalene as a function of the two
PPP model parameters: (a) nearest-neighbor hopping matrix element
$t,$ and (b) on-site repulsion energy, $U$. The calculations for
the purpose were performed using the FCI approach, and both the standard
and the screened parameterization schemes described in Sec. \ref{sec:theory}
were adopted. In order to examine the influence of variation of $t$
on the excitation energy, the values of $U$ was held fixed at 11.13
eV / 8.0 eV for standard/screened parameter calculations, while $t$
was varied in the range depicted in the plot. Similarly, the influence
of variation in $U$ was examined by holding the hopping fixed at
$t=-2.4$ eV , and varying $U$ as per standard/screened parameterization
schemes, in the range specified in the figure. An inspection of Fig.
\ref{fig:Convergence-model-parameters} reveals the following trends
both for the standard and the screened parameter based calculations:
(a) For the fixed value of $U$, with the increasing magnitude of
$t$, the value of the excitation energy increases, and (b) for the
fixed value of $t$, with the increasing value of $U$, the value
of the excitation energy decreases. Furthermore, in ranges of parameters
explored in these calculations, variation of the excitation energy
is almost linear with respect to both $t$ and $U$. Thus, the behavior
the excitation energy of $1^{3}B_{2u}^{+}$ is qualitatively similar
with respect to the model parameters, irrespective of the parameterization
scheme (standard or screened) employed in the calculations.}

\begin{center}
\begin{figure}[H]
\textcolor{red}{\protect}\textcolor{black}{\protect\protect\caption{\label{fig:Convergence-model-parameters}\textcolor{black}{Variation
of the excitation energy of the lowest triplet state, $1^{3}B_{2u}^{+}$,
of naphthalene as a function of the PPP model parameters $t$ (nearest-neighbor
hopping), and $U$ (on-site repulsion energy). Calculations were performed
using the FCI approach using: (a) standard and, (b) screened-type
parameterization.}}
}

\centering{}\textcolor{red}{\includegraphics[angle=-90,width=16cm]{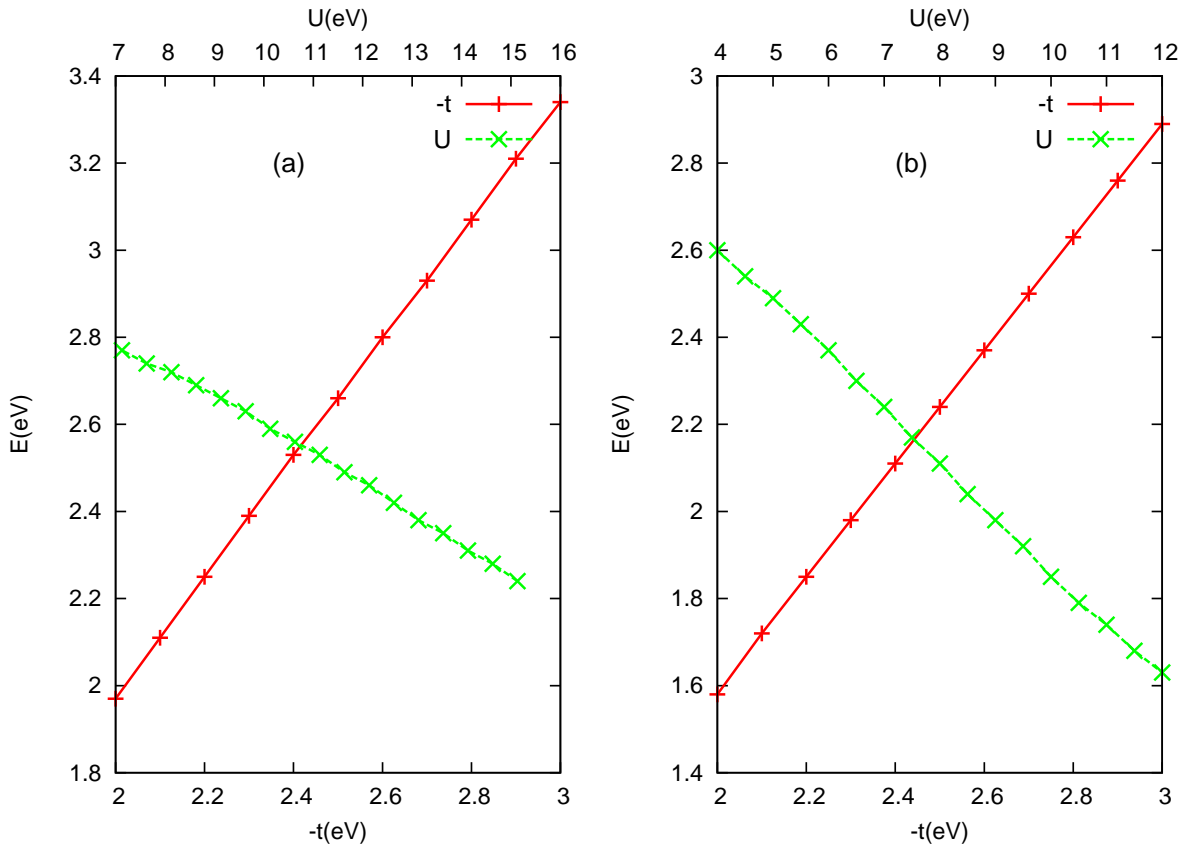}} 
\end{figure}

\par\end{center}

\subsection{\textcolor{black}{Influence of geometry on the triplet optical absorption}}

\textcolor{black}{In this work, consistent with our earlier works,\cite{sony-acene-lo,sony-acene-pa,chakrabortyandshukla2013}
as also the works of Ramasesha and coworkers,\cite{Ramasesha-acene-2002,Ramasesha-instability-12002}
we have used the symmetric ground state geometry for all the oligoacenes,
with all C-C bonds equal to $1.4$ \AA , and all bond angles taken
to be 120$^{o}$. In order to investigate the influence of geometry
on the triplet optical absorption spectra, we performed calculations
on hexacene using a highly non-uniform geometry of $1^{3}B_{2u}^{+}$
state optimized by Houk }\textcolor{black}{\emph{et al}}\textcolor{black}{.,\cite{Houk}
using B3LYP/6-31G{*} calculations. The non-uniformity of this geometry
is obvious from the fact that the smallest C-C bond length in this
structure is $\approx1.38$ Å, while the largest one is close to $1.47$
Å. Similarly, the bond angles in this structure range from 118.57$^{o}$
to 122.68$^{o}$. For this nonuniform geometry, the hopping matrix
elements between nearest-neighbor sites $i$ and $j$, needed for
the PPP calculations, were generated using the exponential formula
$t_{ij}=t_{0}e^{(r_{0}-r_{ij})/\delta}$, where $r_{ij}$ is the bond
distance (in Å) between the sites, $t_{0}=-2.4$ eV, $r_{0}=1.4$
Å, and the decay constant $\delta=0.73$ Å. The value of $\delta$
was chosen so that the formula closely reproduces the hopping matrix
elements for a bond-alternating polyene with short/long bond lengths
$1.35/1.45$ Å. The results of SCI level calculations of the triplet
optical absorption spectrum of hexacene, both for the uniform (symmetric)
and this nonuniform geometry of Houk }\textcolor{black}{\emph{et al}}\textcolor{black}{.\cite{Houk}computed
using the screened parameters of the PPP model, are presented in Fig.
\ref{fig:Convergence-geometry}. From the figure it is obvious that
there are insignificant quantitative differences between the two results,
as far as peak locations are concerned. For example, peaks I and II
of the nonuniform geometry are blue shifted by about 0.1 eV, while
peak III is red shifted by approximately the same amount, as compared
to the corresponding peaks obtained using the uniform geometry. Therefore,
we conclude that the variations in the geometry of the $1^{3}B_{2u}^{+}$
state of the magnitude considered here, lead to small quantitative,
and insignificant qualitative, changes in the triplet optical absorption
spectra of oligoacenes.}

\begin{center}
\begin{figure}[H]
\begin{centering}
\textcolor{red}{\protect}\textcolor{black}{\protect\protect\caption{Triplet optical absorption spectrum of hexacene, with uniform, and
non-uniform geometries\cite{Houk} of the \textcolor{black}{$1^{3}B_{2u}^{+}$}
state, computed at the SCI level using the screened parameters.\label{fig:Convergence-geometry}}
}
\par\end{centering}

\begin{centering}
 
\par\end{centering}

\centering{}\textcolor{red}{\includegraphics[width=14cm]{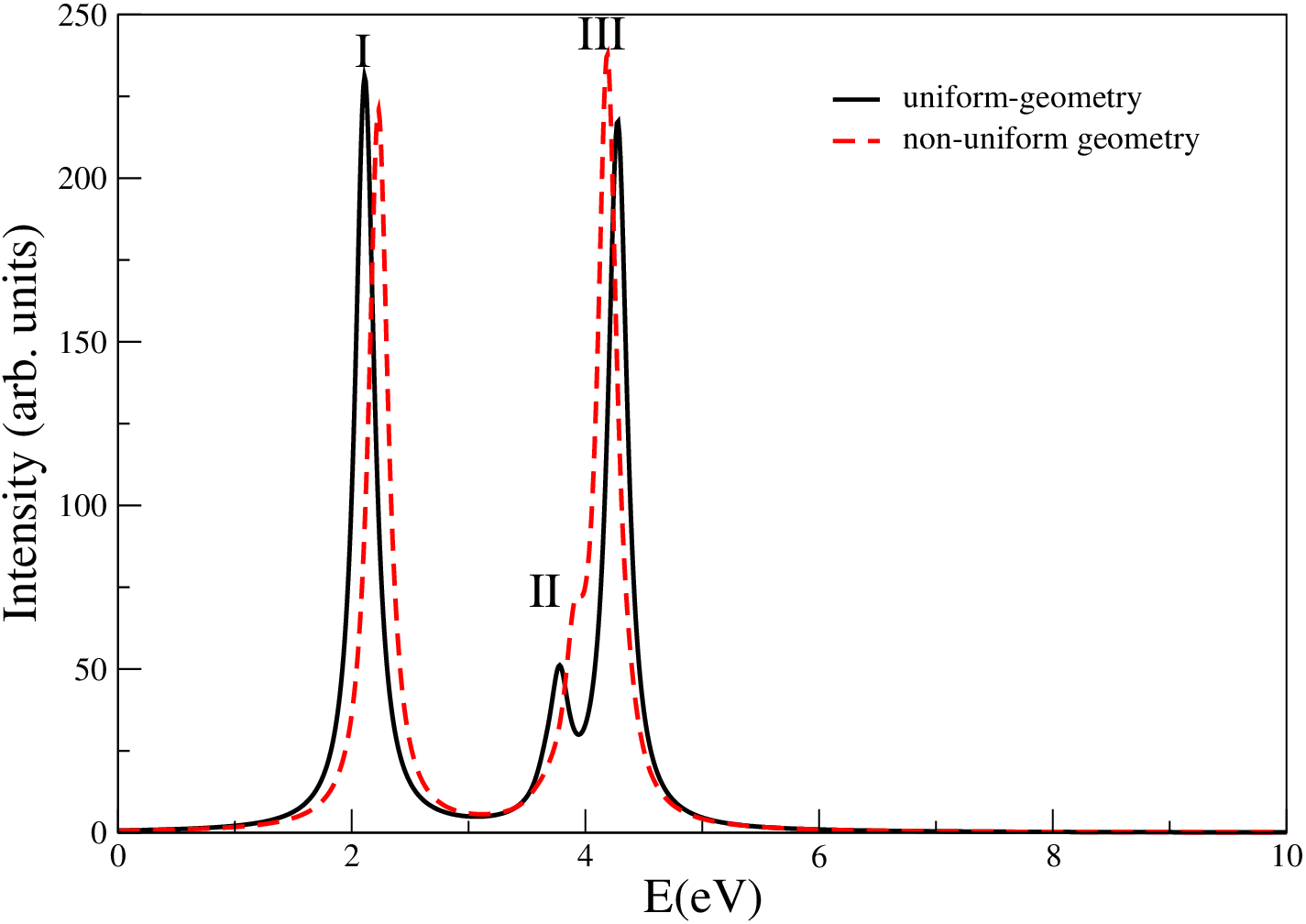}} 
\end{figure}

\par\end{center}

\subsection{\textcolor{black}{Convergence of the MRSDCI excitation energies}}

\textcolor{black}{In this section we demonstrate the convergence of
the excitation energy of the $1^{3}B_{2u}^{+}$ state defined as $E(1^{3}B_{2u}^{+})-E(1^{1}A_{g}^{-})$,
where $E(1^{3}B_{2u}^{+})$, and $E(1^{1}A_{g}^{-})$, respectively,
are the total energies of the $1^{3}B_{2u}^{+}$ state and the ground
state, computed at a similar level of electron-correlation treatment,
employing the MRSDCI approach. The level of correlation treatment
in an MRSDCI calculation can be defined either in terms of the number
of reference configuration one uses in the calculation, or in terms
of the magnitude of the smallest coefficient (henceforth called wave
function cutoff) of the configurations included in the reference list.
It is obvious that smaller the wave function cutoff, larger the number
of configurations in the reference list, and hence larger the total
number of configurations in the MRSDCI expansion. Naturally, for accurate
results, one must use the same cutoff for both the $1^{3}B_{2u}^{+}$
state as well as $1^{1}A_{g}^{-}$ ground state. In Fig. \ref{fig:Convergence-Nref}
we present the results of the MRSDCI calculations of the excitation
energy of the $1^{3}B_{2u}^{+}$ state of anthracene performed using
the screened parameters in the PPP model, and the figure depicts the
behavior of the excitation energy as a function of the wave function
cutoff. We have chosen anthracene, because for this molecule, as reported
in Sec. \ref{sec:results}, FCI results are available, which can be
used to benchmark the accuracy of the MRSDCI calculations. From the
figure it is obvious that with decreasing value of the wave function
cutoff (}\textcolor{black}{\emph{i.e.}}\textcolor{black}{{} increasing
accuracy of the MRSDCI calculation), the excitation energy converges
to a value which is very close to the value 1.48 eV obtained in the
FCI calculations, as reported in Table \ref{tab:Comparison-2-4}}\textcolor{red}{.}

\begin{center}
\begin{figure}[H]
\textcolor{red}{\protect}\textcolor{black}{\protect\protect\caption{\label{fig:Convergence-Nref}Convergence of the excitation energy
of the lowest triplet state $1^{3}B_{2u}^{+}$ of anthracene with
respect to the MRSDCI wave function cutoff (see text for an explanation).
All calculations were performed using the screened parameters in the
PPP model, and the FCI value of the excitation energy is 1.480 eV
(\emph{cf}. Table \ref{tab:Comparison-2-4}). }
}

\centering{}\textcolor{red}{\includegraphics[angle=-90,width=14cm]{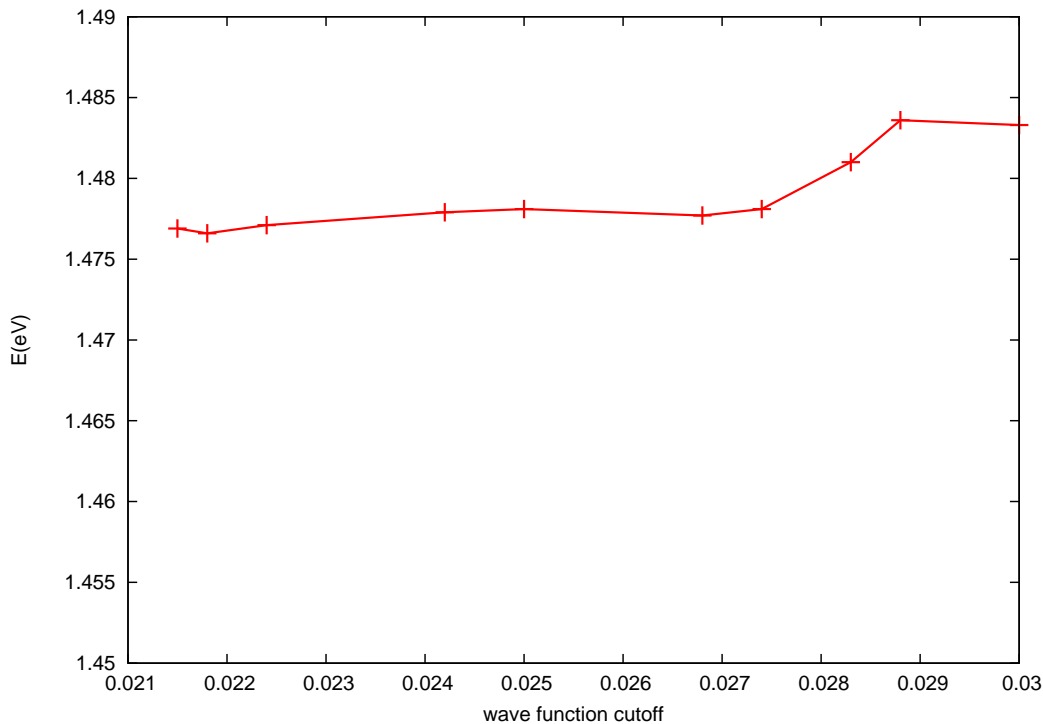}} 
\end{figure}

\par\end{center}

\section{Calculated Energies, Wave Functions, and Transition Dipole Moments
of the Excited States Contributing to the Triplet Absorption Spectra
\label{sec:appendixB}}

In this section, we present Tables \ref{tab:acene2-pa-std}---\ref{tab:acene7-pa-scr}
which contain the excitation energies, transition dipole moments,
and wave functions of the triplet excited states, which contribute
to the calculated triplet absorption spectra of oligoacenes ranging
from naphthalene to heptacene, presented in Figs. \ref{fig:acene2_pa}---\ref{fig:acene7_pa}
of the main text.

\begin{center}
\begin{table}[H]
\protect\protect\protect\caption{\textcolor{black}{Properties of the excited states leading to various
peaks in the triplet absorption spectrum of naphthalene computed using
the FCI method, and the standard parameters in the PPP model Hamiltonian.
The table contains excitation energies (with respect to the $1^{1}A_{g}^{-}$,
ground state), many-particle wave functions, and the transition dipole
matrix elements of various states, connecting them to the lowest triplet
state $1^{3}B_{2u}^{+}$. Below, `$+c.c.$' indicates that the coefficient
of the electron-hole conjugate of a given configuration has the same
sign, while `$-c.c.$' implies that they have opposite signs. DF denotes
that the concerned excited state is dipole forbidden. \label{tab:acene2-pa-std}}}

\vspace{0.25cm}

\centering{}%
\begin{tabular}{|c|c|c|c|c|}
\hline 
Peak  & State  & E (eV)  & Transition  & dominant contributing configurations\tabularnewline
 &  &  & Dipole (\AA )  & \tabularnewline
DF  & $1^{3}B_{1g}^{+}$  & 3.74  & 0.0  & $\mid H\rightarrow L+2\left\rangle \right.-c.c.(0.6421)$\tabularnewline
 &  &  &  & $\mid H-1\rightarrow L+3\left\rangle \right.+c.c.(0.1200)$\tabularnewline
DF  & $1^{3}A{}_{g}^{+}$  & 4.59  & 0.0  & $\mid H\rightarrow L+3\left\rangle \right.+c.c.(0.5355)$\tabularnewline
 &  &  &  & $\mid H-1\rightarrow L+2\left\rangle \right.-c.c.(0.3475)$\tabularnewline
\hline 
I  & $1^{3}B_{1g}^{-}$  & 6.13  & 0.739  & $\mid H\rightarrow L+2\left\rangle \right.+c.c.(0.6225)$\tabularnewline
 &  &  &  & $\mid H\rightarrow L;H\rightarrow L+1\left\rangle \right.-c.c.(0.2044)$\tabularnewline
\hline 
II  & $1^{3}A{}_{g}^{-}$  & 6.29  & 0.225  & $\mid H-1\rightarrow L+2\left\rangle \right.+c.c.(0.5890)$\tabularnewline
 &  &  &  & $\mid H\rightarrow L+3\left\rangle \right.-c.c.(0.2993)$\tabularnewline
\hline 
III  & $2^{3}A{}_{g}^{-}$  & 7.3  & 0.590  & $\mid H\rightarrow L+3\left\rangle \right.-c.c.(0.5469)$\tabularnewline
 &  &  &  & $\mid H-1\rightarrow L+2\left\rangle \right.+c.c.(02679)$\tabularnewline
\hline 
IV  & $2^{3}B_{1g}^{-}$  & 7.9  & 0.608  & $\mid H-1\rightarrow L+3\left\rangle \right.-c.c.(0.5168)$\tabularnewline
 &  &  &  & $\mid H\rightarrow L;H\rightarrow L+1\left\rangle \right.-c.c.(0.2928)$\tabularnewline
\hline 
V  & $3^{3}B_{1g}^{-}$  & 8.61  & 1.142  & $\mid H\rightarrow L;H\rightarrow L+1\left\rangle \right.-c.c.(0.4939)$\tabularnewline
 &  &  &  & $\mid H-1\rightarrow L+3\left\rangle \right.-c.c.(0.2839)$\tabularnewline
\hline 
\end{tabular}
\end{table}

\par\end{center}

\begin{center}
\begin{table}
\protect\protect\protect\caption{\textcolor{black}{Properties of the excited states leading to various
peaks in the triplet absorption spectrum of naphthalene computed using
the FCI method, and the screened parameters in the PPP model Hamiltonian.
The rest of the information is same as in the caption of Table \ref{tab:acene2-pa-std}.\label{tab:acene2-pa-scr}}}

\vspace{0.25cm}

\centering{}%
\begin{tabular}{|c|c|c|c|c|}
\hline 
Peak  & State  & E (eV)  & Transition  & dominant contributing configurations\tabularnewline
 &  &  & Dipole (\AA )  & \tabularnewline
DF  & $1^{3}B_{1g}^{+}$  & 3.27  & 0.0  & $\mid H\rightarrow L+2\left\rangle \right.+c.c.(0.6148)$\tabularnewline
 &  &  &  & $\mid H-1\rightarrow L+3\left\rangle \right.-c.c.(0.1463)$\tabularnewline
DF  & $1^{3}A{}_{g}^{+}$  & 4.08  & 0.0  & $\mid H\rightarrow L+3\left\rangle \right.+c.c.(0.5349)$\tabularnewline
 &  &  &  & $\mid H-1\rightarrow L+2\left\rangle \right.-c.c.(0.2822)$\tabularnewline
\hline 
I  & $1^{3}B_{1g}^{-}$  & 5.61  & 0.940  & $\mid H\rightarrow L+2\left\rangle \right.-c.c.(0.6447)$\tabularnewline
 &  &  &  & $\mid H\rightarrow L+2;H-3\rightarrow L\left\rangle \right.c.c.(0.1033)$\tabularnewline
\hline 
II  & $1^{3}A{}_{g}^{-}$  & 6.23  & 0.182  & $\mid H-1\rightarrow L+2\left\rangle \right.+c.c.(0.5731)$\tabularnewline
 &  &  &  & $\mid H\rightarrow L+3\left\rangle \right.-c.c.(0.3015)$\tabularnewline
\hline 
III  & $2^{3}A{}_{g}^{-}$  & 6.48  & 0.579  & $\mid H\rightarrow L+3\left\rangle \right.-c.c.(0.5535)$\tabularnewline
 &  &  &  & $\mid H-1\rightarrow L+2\left\rangle \right.+c.c.(0.2966)$\tabularnewline
\hline 
IV  & $2^{3}B_{1g}^{-}$  & 7.16  & 0.234  & $\mid H-1\rightarrow L+3\left\rangle \right.+c.c.(0.5873)$\tabularnewline
 &  &  &  & $\mid H\rightarrow L;H\rightarrow L+1\left\rangle \right.+c.c.(0.2220)$\tabularnewline
\hline 
V  & $3^{3}B_{1g}^{-}$  & 7.75  & 1.109  & $\mid H\rightarrow L;H\rightarrow L+1\left\rangle \right.+c.c.(0.5609)$\tabularnewline
 &  &  &  & $\mid H-1\rightarrow L+3\left\rangle \right.+c.c.(0.1903)$\tabularnewline
\hline 
VI  & $3^{3}A{}_{g}^{-}$  & 8.47  & 0.466  & $\mid H\rightarrow L;H-1\rightarrow L+1\left\rangle \right.(0.8268)$\tabularnewline
 &  &  &  & $\mid H\rightarrow L+3;H-3\rightarrow L\left\rangle \right.(0.2105)$\tabularnewline
\hline 
\end{tabular}
\end{table}

\par\end{center}

\begin{center}
\begin{table}[H]
\protect\protect\protect\caption{\textcolor{black}{Properties of the excited states leading to various
peaks in the triplet absorption spectrum of anthracene computed using
the FCI method, and the standard parameters in the PPP model Hamiltonian.
The rest of the information is same as in the caption of Table \ref{tab:acene2-pa-std}.\label{tab:acene3-pa-std}}}

\vspace{0.25cm}

\centering{}%
\begin{tabular}{|c|c|c|c|c|}
\hline 
Peak  & State  & E (eV)  & Transition  & Dominant Contributing Configurations\tabularnewline
 &  &  & Dipole (\AA )  & \tabularnewline
DF  & $1^{3}B_{1g}^{+}$  & 2.95  & 0.0  & $\mid H\rightarrow L+2\left\rangle \right.+c.c.(0.5992)$\tabularnewline
 &  &  &  & $\mid H-2\rightarrow L+4\left\rangle \right.+c.c.(0.1889)$\tabularnewline
DF  & $1^{3}A{}_{g}^{+}$  & 4.04  & 0.0  & $\mid H\rightarrow L+3\left\rangle \right.-c.c.(0.5106)$\tabularnewline
 &  &  &  & $\mid H-1\rightarrow L+2\left\rangle \right.-c.c.(0.3024)$\tabularnewline
\hline 
I  & $1^{3}B_{1g}^{-}$  & 5.03  & 1.132  & $\mid H\rightarrow L+2\left\rangle \right.-c.c.(0.5986)$\tabularnewline
 &  &  &  & $\mid H\rightarrow L;H\rightarrow L+1\left\rangle \right.+c.c.(0.1844)$\tabularnewline
\hline 
II  & $1^{3}A{}_{g}^{-}$  & 5.57  & 0.356  & $\mid H-1\rightarrow L+2\left\rangle \right.+c.c.(0.4912)$\tabularnewline
 &  &  &  & $\mid H\rightarrow L+3\left\rangle \right.+c.c.(0.3989)$\tabularnewline
\hline 
III  & $2^{3}A{}_{g}^{-}$  & 6.45  & 0.482  & $\mid H\rightarrow L+3\left\rangle \right.+c.c.(0.4523)$\tabularnewline
 &  &  &  & $\mid H-1\rightarrow L+2\left\rangle \right.+c.c.(0.3453)$\tabularnewline
\hline 
IV  & $2^{3}B_{1g}^{-}$  & 6.95  & 1.568  & $\mid H\rightarrow L;H\rightarrow L+1\left\rangle \right.+c.c.(0.5360)$\tabularnewline
 &  &  &  & $\mid H-1\rightarrow L+3\left\rangle \right.-c.c.(0.1850)$\tabularnewline
\hline 
V  & $3^{3}B_{1g}^{-}$  & 7.58  & 0.309  & $\mid H-1\rightarrow L+3\left\rangle \right.-c.c.(0.3752)$\tabularnewline
 &  &  &  & $\mid H-2\rightarrow L+4\left\rangle \right.-c.c.(0.3364)$\tabularnewline
 & $4^{3}A{}_{g}^{-}$  & 7.54  & 0.236  & $\mid H\rightarrow L;H-2\rightarrow L+2\left\rangle \right.(0.5992)$\tabularnewline
 &  &  &  & $\mid H\rightarrow L;H-1\rightarrow L+1\left\rangle \right.(0.5288)$\tabularnewline
\hline 
VI  & $4^{3}B_{1g}^{-}$  & 7.83  & 0.356  & $\mid H-2\rightarrow L+4\left\rangle \right.-c.c.(0.4031)$\tabularnewline
 &  &  &  & $\mid H-1\rightarrow L+3\left\rangle \right.-c.c.(0.3235)$\tabularnewline
\hline 
\end{tabular}
\end{table}

\par\end{center}

\begin{center}
\begin{table}[H]
\protect\protect\protect\caption{\textcolor{black}{Properties of the excited states leading to various
peaks in the triplet absorption spectrum of anthracene computed using
the FCI method, and the screened parameters in the PPP model Hamiltonian.
The rest of the information is same as in the caption of Table \ref{tab:acene2-pa-std}.
\label{tab:acene3-pa-scr}}}

\vspace{0.25cm}

\centering{}%
\begin{tabular}{|c|c|c|c|c|}
\hline 
Peak  & State  & E (eV)  & Transition  & Dominant Contributing Configurations\tabularnewline
 &  &  & Dipole (\AA )  & \tabularnewline
DF  & $1^{3}B_{1g}^{+}$  & 2.58  & 0.0  & $\mid H\rightarrow L+2\left\rangle \right.-c.c.(0.5735)$\tabularnewline
 &  &  &  & $\mid H-2\rightarrow L+4\left\rangle \right.+c.c.(0.1776)$\tabularnewline
DF  & $1^{3}A{}_{g}^{+}$  & 3.59  & 0.0  & $\mid H\rightarrow L+3\left\rangle \right.+c.c.(0.5042)$\tabularnewline
 &  &  &  & $\mid H-1\rightarrow L+2\left\rangle \right.+c.c.(0.2449)$\tabularnewline
\hline 
I  & $1^{3}B_{1g}^{-}$  & 4.57  & 1.394  & $\mid H\rightarrow L+2\left\rangle \right.+c.c.(0.6219)$\tabularnewline
 &  &  &  & $\mid H\rightarrow L;H-1\rightarrow L\left\rangle \right.+c.c.(0.0854)$\tabularnewline
\hline 
II  & $1^{3}A{}_{g}^{-}$  & 5.44  & 0.521  & $\mid H\rightarrow L+3\left\rangle \right.-c.c.(0.5463)$\tabularnewline
 &  &  &  & $\mid H-1\rightarrow L+2\left\rangle \right.-c.c.(0.2938)$\tabularnewline
\hline 
III  & $2^{3}A{}_{g}^{-}$  & 5.72  & 0.370  & $\mid H-1\rightarrow L+2\left\rangle \right.-c.c.(0.5371)$\tabularnewline
 &  &  &  & $\mid H\rightarrow L+3\left\rangle \right.-c.c.(0.2765)$\tabularnewline
\hline 
IV  & $2^{3}B_{1g}^{-}$  & 6.29  & 1.335  & $\mid H\rightarrow L;H\rightarrow L+1\left\rangle \right.+c.c.(0.5577)$\tabularnewline
 &  &  &  & $\mid H-1\rightarrow L+3\left\rangle \right.+c.c.(0.2127)$\tabularnewline
\hline 
V  & $3^{3}B_{1g}^{-}$  & 6.78  & 0.544  & $\mid H-1\rightarrow L+3\left\rangle \right.+c.c.(0.5194)$\tabularnewline
 &  &  &  & $\mid H\rightarrow L;H\rightarrow L+1\left\rangle \right.+c.c.(0.1673)$\tabularnewline
\hline 
VI  & $4^{3}A{}_{g}^{-}$  & 7.36  & 0.530  & $\mid H\rightarrow L;H-1\rightarrow L+1\left\rangle \right.(0.5641)$\tabularnewline
 &  &  &  & $\mid H\rightarrow L;H-2\rightarrow L+2\left\rangle \right.(0.4119)$\tabularnewline
 & $5^{3}A{}_{g}^{-}$  & 7.40  & 0.387  & $\mid H\rightarrow L;H-2\rightarrow L+2\left\rangle \right.(0.6385)$\tabularnewline
 &  &  &  & $\mid H\rightarrow L+6\left\rangle \right.-c.c.(0.3256)$\tabularnewline
\hline 
VII  & $6^{3}A{}_{g}^{-}$  & 7.64  & 0.652  & $\mid H\rightarrow L;H-1\rightarrow L+1\left\rangle \right.(0.7058)$\tabularnewline
 &  &  &  & $\mid H-3\rightarrow L+4\left\rangle \right.+c.c.(0.2388)$\tabularnewline
\hline 
\end{tabular}
\end{table}

\par\end{center}

\begin{center}
\begin{table}[H]
\protect\protect\protect\caption{\textcolor{black}{Properties of the excited states leading to various
peaks in the triplet absorption spectrum of tetracene computed using
the MRSDCI method, and the standard parameters in the PPP model Hamiltonian.
The rest of the information is same as in the caption of Table \ref{tab:acene2-pa-std}.\label{tab:acene4-pa-std}}}

\vspace{0.25cm}

\centering{}%
\begin{tabular}{|c|c|c|c|c|}
\hline 
Peak  & State  & E (eV)  & Transition  & Dominant Contributing Configurations\tabularnewline
 &  &  & Dipole (\AA )  & \tabularnewline
DF  & $1^{3}B_{1g}^{+}$  & 2.34  & 0.0  & $\mid H\rightarrow L+1\left\rangle \right.-c.c.(0.5959)$\tabularnewline
 &  &  &  & $\mid H-1\rightarrow L+3\left\rangle \right.+c.c.(0.1194)$\tabularnewline
DF  & $1^{3}A{}_{g}^{+}$  & 3.80  & 0.0  & $\mid H\rightarrow L+4\left\rangle \right.-c.c.(0.4892)$\tabularnewline
 &  &  &  & $\mid H-1\rightarrow L+2\left\rangle \right.+c.c.(0.3127)$\tabularnewline
\hline 
I  & $1^{3}B_{1g}^{-}$  & 4.30  & 1.501  & $\mid H\rightarrow L+1\left\rangle \right.+c.c.(0.5876)$\tabularnewline
 &  &  &  & $\mid H\rightarrow L;H\rightarrow L+2\left\rangle \right.+c.c.(0.1630)$\tabularnewline
\hline 
II  & $1^{3}A{}_{g}^{-}$  & 5.07  & 0.356  & $\mid H-1\rightarrow L+2\left\rangle \right.-c.c.(0.4523)$\tabularnewline
 &  &  &  & $\mid H\rightarrow L+4\left\rangle \right.+c.c.(0.4234)$\tabularnewline
\hline 
III  & $3^{3}B_{1g}^{-}$  & 6.23  & 1.865  & $\mid H\rightarrow L;H\rightarrow L+2\left\rangle \right.+c.c.(0.5446)$\tabularnewline
 &  &  &  & $\mid H-1\rightarrow L;H\rightarrow L+4\left\rangle \right.c.c.(0.1402)$\tabularnewline
\hline 
IV  & $3^{3}A{}_{g}^{-}$  & 6.59  & 0.346  & $\mid H\rightarrow L;H-1\rightarrow L+1\left\rangle \right.(0.4758)$\tabularnewline
 &  &  &  & $\mid H\rightarrow L;H\rightarrow L+3\left\rangle \right.+c.c.(0.4152)$\tabularnewline
 & $4^{3}A{}_{g}^{-}$  & 6.64  & 0.654  & $\mid H\rightarrow L;H-1\rightarrow L+1\left\rangle \right.(0.6786)$\tabularnewline
 &  &  &  & $\mid H\rightarrow L+4\left\rangle \right.+c.c.(0.1947)$\tabularnewline
\hline 
V  & $6^{3}A{}_{g}^{-}$  & 7.39  & 0.239  & $\mid H\rightarrow L;H-1\rightarrow L+1\left\rangle \right.(0.4877)$\tabularnewline
 &  &  &  & $\mid H\rightarrow L;H-2\rightarrow L+2\left\rangle \right.(0.3882)$\tabularnewline
\hline 
\end{tabular}
\end{table}

\par\end{center}

\begin{center}
\begin{table}[H]
\protect\protect\protect\caption{\textcolor{black}{Properties of the excited states leading to various
peaks in the triplet absorption spectrum of tetracene computed using
the MRSDCI method, and the screened parameters in the PPP model Hamiltonian.
The rest of the information is same as in the caption of Table \ref{tab:acene2-pa-std}.
\label{tab:acene4-pa-scr}}}

\vspace{0.25cm}

\centering{}%
\begin{tabular}{|c|c|c|c|c|}
\hline 
Peak  & State  & E (eV)  & Transition  & Dominant Contributing Configurations\tabularnewline
 &  &  & Dipole (\AA )  & \tabularnewline
DF  & $1^{3}B_{1g}^{+}$  & 2.15  & 0.0  & $\mid H\rightarrow L+1\left\rangle \right.+c.c.(0.5828)$\tabularnewline
 &  &  &  & $\mid H-1\rightarrow L+3\left\rangle \right.+c.c.(0.1199)$\tabularnewline
DF  & $1^{3}A{}_{g}^{+}$  & 3.53  & 0.0  & $\mid H\rightarrow L+4\left\rangle \right.-c.c.(0.5092)$\tabularnewline
 &  &  &  & $\mid H-1\rightarrow L+2\left\rangle \right.-c.c.(0.2542)$\tabularnewline
\hline 
I  & $1^{3}B_{1g}^{-}$  & 3.90  & 1.824  & $\mid H\rightarrow L+1\left\rangle \right.-c.c.(0.6176)$\tabularnewline
 &  &  &  & $\mid H\rightarrow L+1;H-1\rightarrow L+2\left\rangle \right.c.c.(0.0735)$\tabularnewline
\hline 
II  & $1^{3}A{}_{g}^{-}$  & 5.02  & 0.524  & $\mid H\rightarrow L+4\left\rangle \right.+c.c.(0.5638)$\tabularnewline
 &  &  &  & $\mid H-1\rightarrow L+2\left\rangle \right.+c.c.(0.2617)$\tabularnewline
\hline 
III  & $3^{3}B_{1g}^{-}$  & 5.78  & 1.648  & $\mid H\rightarrow L;H\rightarrow L+2\left\rangle \right.+c.c.(0.5793)$\tabularnewline
 &  &  &  & $\mid H-1\rightarrow L;H-2\rightarrow L+1\left\rangle \right.c.c.(0.1355)$\tabularnewline
\hline 
IV  & $4^{3}B_{1g}^{-}$  & 6.28  & 0.203  & $\mid H-1\rightarrow L+3\left\rangle \right.-c.c.(0.5696)$\tabularnewline
 &  &  &  & $\mid H\rightarrow L;H\rightarrow L;H-1\rightarrow L+3\left\rangle \right.-c.c.(0.1273)$\tabularnewline
 & $3^{3}A{}_{g}^{-}$  & 6.31  & 0.779  & $\mid H\rightarrow L;H-1\rightarrow L+1\left\rangle \right.(0.8320)$\tabularnewline
 &  &  &  & $\mid H\rightarrow L+1;H-1\rightarrow L+3\left\rangle \right.-c.c.(0.1013)$\tabularnewline
\hline 
V  & $7^{3}B_{1g}^{-}$  & 7.33  & 0.503  & $\mid H\rightarrow L;H-2\rightarrow L+3\left\rangle \right.-c.c.(0.3526)$\tabularnewline
 &  &  &  & $\mid H\rightarrow L+1;H-4\rightarrow L\left\rangle \right.-c.c.(0.3303)$\tabularnewline
 & $8^{3}A{}_{g}^{-}$  & 7.42  & 0.785  & $\mid H\rightarrow L;H-2\rightarrow L+2\left\rangle \right.(0.7912)$\tabularnewline
 &  &  &  & $\mid H-1\rightarrow L+2;H-2\rightarrow L+1\left\rangle \right.(0.1316)$\tabularnewline
\hline 
\end{tabular}
\end{table}

\par\end{center}

\begin{center}
\begin{table}[H]
\protect\protect\protect\caption{\textcolor{black}{Properties of the excited states leading to various
peaks in the triplet absorption spectrum of pentacene computed using
the MRSDCI method, and the standard parameters in the PPP model Hamiltonian.
The rest of the information is same as in the caption of Table \ref{tab:acene2-pa-std}.
\label{tab:acene5-pa-std}}}

\vspace{0.25cm}

\centering{}%
\begin{tabular}{|c|c|c|c|c|}
\hline 
Peak  & State  & E (eV)  & Transition  & Dominant Contributing Configurations\tabularnewline
 &  &  & Dipole (\AA )  & \tabularnewline
DF  & $1^{3}B_{1g}^{+}$  & 1.95  & 0.0  & $\mid H\rightarrow L+1\left\rangle \right.-c.c.(0.5890)$\tabularnewline
 &  &  &  & $\mid H-1\rightarrow L+3\left\rangle \right.-c.c.(0.1143)$\tabularnewline
DF  & $1^{3}A{}_{g}^{+}$  & 3.66  & 0.0  & $\mid H\rightarrow L+4\left\rangle \right.-c.c.(0.4589)$\tabularnewline
 &  &  &  & $\mid H-1\rightarrow L+2\left\rangle \right.+c.c.(0.3251)$\tabularnewline
\hline 
I  & $1^{3}B_{1g}^{-}$  & 3.86  & 1.842  & $\mid H\rightarrow L+1\left\rangle \right.+c.c.(0.5785)$\tabularnewline
 &  &  &  & $\mid H\rightarrow L;H\rightarrow L+2\left\rangle \right.+c.c.(0.1413)$\tabularnewline
\hline 
II  & $1^{3}A{}_{g}^{-}$  & 4.76  & 0.303  & $\mid H-1\rightarrow L+2\left\rangle \right.-c.c.(0.4328)$\tabularnewline
 &  &  &  & $\mid H\rightarrow L+4\left\rangle \right.+c.c.(0.4179)$\tabularnewline
\hline 
III  & $3^{3}B_{1g}^{-}$  & 5.84  & 2.02  & $\mid H\rightarrow L;H\rightarrow L+2\left\rangle \right.+c.c.(0.5287)$\tabularnewline
 &  &  &  & $\mid H-1\rightarrow L;H\rightarrow L+4\left\rangle \right.c.c.(0.1615)$\tabularnewline
\hline 
IV  & $3^{3}A{}_{g}^{-}$  & 6.09  & 0.373  & $\mid H\rightarrow L;H-1\rightarrow L+1\left\rangle \right.(0.5984)$\tabularnewline
 &  &  &  & $\mid H\rightarrow L;H\rightarrow L+3\left\rangle \right.-c.c.(0.3464)$\tabularnewline
 & $4^{3}A{}_{g}^{-}$  & 6.15  & 0.706  & $\mid H\rightarrow L;H-1\rightarrow L+1\left\rangle \right.(0.3977)$\tabularnewline
 &  &  &  & $\mid H\rightarrow L+4\left\rangle \right.+c.c.(0.3550)$\tabularnewline
\hline 
V  & $6^{3}B_{1g}^{-}$  & 7.15  & 0.669  & $\mid H\rightarrow L+1;H-4\rightarrow L\left\rangle \right.-c.c.(0.3978)$\tabularnewline
 &  &  &  & $\mid H-1\rightarrow L+2;H\rightarrow L+1\left\rangle \right.-c.c.(0.2217)$\tabularnewline
\hline 
\end{tabular}
\end{table}

\par\end{center}

\begin{center}
\begin{table}[H]
\protect\protect\protect\caption{\textcolor{black}{Properties of the excited states leading to various
peaks in the triplet absorption spectrum of pentacene computed using
the MRSDCI method, and the screened parameters in the PPP model Hamiltonian.
The rest of the information is same as in the caption of Table \ref{tab:acene2-pa-std}.\label{tab:acene5-pa-scr}}}

\vspace{0.25cm}

\centering{}%
\begin{tabular}{|c|c|c|c|c|}
\hline 
Peak  & State  & E (eV)  & Transition  & Dominant Contributing Configurations\tabularnewline
 &  &  & Dipole (\AA )  & \tabularnewline
DF  & $1^{3}B_{1g}^{+}$  & 1.83  & 0.0  & $\mid H\rightarrow L+1\left\rangle \right.-c.c.(0.5815)$\tabularnewline
 &  &  &  & $\mid H-1\rightarrow L+3\left\rangle \right.-c.c.(0.1207)$\tabularnewline
DF  & $1^{3}A{}_{g}^{+}$  & 3.37  & 0.0  & $\mid H\rightarrow L+4\left\rangle \right.+c.c.(0.4773)$\tabularnewline
 &  &  &  & $\mid H-1\rightarrow L+2\left\rangle \right.-c.c.(0.2856)$\tabularnewline
\hline 
I  & $1^{3}B_{1g}^{-}$  & 3.47  & 2.208  & $\mid H\rightarrow L+1\left\rangle \right.+c.c.(0.6091)$\tabularnewline
 &  &  &  & $\mid H-1\rightarrow L+3\left\rangle \right.+c.c.(0.0766)$\tabularnewline
\hline 
II  & $1^{3}A{}_{g}^{-}$  & 4.72  & 0.396  & $\mid H\rightarrow L+4\left\rangle \right.-c.c.(0.5163)$\tabularnewline
 &  &  &  & $\mid H-1\rightarrow L+2\left\rangle \right.+c.c.(0.3126)$\tabularnewline
\hline 
III  & $3^{3}B_{1g}^{-}$  & 5.51  & 1.816  & $\mid H\rightarrow L;H\rightarrow L+2\left\rangle \right.+c.c.(0.5701)$\tabularnewline
 &  &  &  & $\mid H-1\rightarrow L;H-2\rightarrow L+1\left\rangle \right.c.c.(0.1440)$\tabularnewline
\hline 
IV  & $5^{3}B_{1g}^{-}$  & 6.78  & 0.743  & $\mid H\rightarrow L+1;H-4\rightarrow L\left\rangle \right.-c.c.(0.3970)$\tabularnewline
 &  &  &  & $\mid H\rightarrow L+1;H-1\rightarrow L+2\left\rangle \right.-c.c.(0.2538)$\tabularnewline
\hline 
V  & $11^{3}A{}_{g}^{-}$  & 7.33  & 0.805  & $\mid H\rightarrow L;H-2\rightarrow L+2\left\rangle \right.(0.7543)$\tabularnewline
 &  &  &  & $\mid H-1\rightarrow L+7\left\rangle \right.+c.c.(0.1390)$\tabularnewline
 & $12^{3}A{}_{g}^{-}$  & 7.35  & 0.334  & $\mid H\rightarrow L+3;H-1\rightarrow L+1\left\rangle \right.-c.c.(0.3503)$\tabularnewline
 &  &  &  & $\mid H\rightarrow L;H-1\rightarrow L+5\left\rangle \right.-c.c.(0.3135)$\tabularnewline
\hline 
\end{tabular}
\end{table}

\par\end{center}

\begin{center}
\begin{table}[H]
\protect\protect\protect\caption{\textcolor{black}{Properties of the excited states leading to various
peaks in the triplet absorption spectrum of hexacene computed using
the MRSDCI method, and the standard parameters in the PPP model Hamiltonian.
The rest of the information is same as in the caption of Table \ref{tab:acene2-pa-std}.}\label{tab:acene6-pa-std}}

\vspace{0.25cm}

\centering{}%
\begin{tabular}{|c|c|c|c|c|}
\hline 
Peak  & State  & E (eV)  & Transition  & Dominant Contributing Configurations\tabularnewline
 &  &  & Dipole (\AA )  & \tabularnewline
DF  & $1^{3}B_{1g}^{+}$  & 1.67  & 0.0  & $\mid H\rightarrow L+1\left\rangle \right.+c.c.(0.5841)$\tabularnewline
 &  &  &  & $\mid H-1\rightarrow L+2\left\rangle \right.+c.c.(0.1239)$\tabularnewline
DF  & $1^{3}A{}_{g}^{+}$  & 3.50  & 0.0  & $\mid H\rightarrow L+4\left\rangle \right.-c.c.(0.4127)$\tabularnewline
 &  &  &  & $\mid H-1\rightarrow L+3\left\rangle \right.+c.c.(0.3077)$\tabularnewline
\hline 
I  & $1^{3}B_{1g}^{-}$  & 3.57  & 2.153  & $\mid H\rightarrow L+1\left\rangle \right.-c.c.(0.5705)$\tabularnewline
 &  &  &  & $\mid H-1\rightarrow L+2\left\rangle \right.-c.c.(0.1471)$\tabularnewline
\hline 
 & $1^{3}A{}_{g}^{-}$  & 4.61  & 0.218  & $\mid H-1\rightarrow L+3\left\rangle \right.+c.c.(0.4363)$\tabularnewline
 &  &  &  & $\mid H\rightarrow L+4\left\rangle \right.+c.c.(0.4341)$\tabularnewline
\hline 
II  & $2^{3}A{}_{g}^{-}$  & 5.15  & 0.647  & $\mid H\rightarrow L;H-1\rightarrow L+1\left\rangle \right.(0.8093)$\tabularnewline
 &  &  &  & $\mid H\rightarrow L;H-3\rightarrow L+3\left\rangle \right.(0.1337)$\tabularnewline
\hline 
III  & $3^{3}B_{1g}^{-}$  & 5.62  & 2.119  & $\mid H\rightarrow L;H\rightarrow L+3\left\rangle \right.-c.c.(0.5159)$\tabularnewline
 &  &  &  & $\mid H\rightarrow L+1;H-4\rightarrow L\left\rangle \right.c.c.(0.1642)$\tabularnewline
\hline 
IV  & $4^{3}A{}_{g}^{-}$  & 5.98  & 0.482  & $\mid H\rightarrow L+4\left\rangle \right.+c.c.(0.3650)$\tabularnewline
 &  &  &  & $\mid H-1\rightarrow L+3\left\rangle \right.-c.c.(0.3138)$\tabularnewline
 & $4^{3}B_{1g}^{-}$  & 5.98  & 0.380  & $\mid H-1\rightarrow L+2\left\rangle \right.-c.c.(0.4471)$\tabularnewline
 &  &  &  & $\mid H\rightarrow L+1;H-1\rightarrow L+3\left\rangle \right.-c.c.(0.2035)$\tabularnewline
\hline 
V  & $7^{3}B_{1g}^{-}$  & 6.82  & 1.056  & $\mid H\rightarrow L+1;H-4\rightarrow L\left\rangle \right.-c.c.(0.4144)$\tabularnewline
 &  &  &  & $\mid H\rightarrow L+1;H-1\rightarrow L+3\left\rangle \right.-c.c.(0.2725)$\tabularnewline
\hline 
\end{tabular}
\end{table}

\par\end{center}

\begin{center}
\begin{table}[H]
\protect\protect\protect\caption{\textcolor{black}{Properties of the excited states leading to various
peaks in the triplet absorption spectrum of hexacene computed using
the MRSDCI method, and the screened parameters in the PPP model Hamiltonian.
The rest of the information is same as in the caption of Table \ref{tab:acene2-pa-std}.\label{tab:acene6-pa-scr}}}

\vspace{0.25cm}

\centering{}%
\begin{tabular}{|c|c|c|c|c|}
\hline 
Peak  & State  & E (eV)  & Transition  & Dominant Contributing Configurations\tabularnewline
 &  &  & Dipole (\AA )  & \tabularnewline
DF  & $1^{3}B_{1g}^{+}$  & 1.60  & 0.0  & $\mid H\rightarrow L+1\left\rangle \right.+c.c.(0.5799)$\tabularnewline
 &  &  &  & $\mid H-1\rightarrow L+2\left\rangle \right.+c.c.(0.1197)$\tabularnewline
DF  & $1^{3}A{}_{g}^{+}$  & 3.10  & 0.0  & $\mid H\rightarrow L+4\left\rangle \right.+c.c.(0.4573)$\tabularnewline
 &  &  &  & $\mid H-1\rightarrow L+3\left\rangle \right.+c.c.(0.2732)$\tabularnewline
\hline 
I  & $1^{3}B_{1g}^{-}$  & 3.18  & 2.485  & $\mid H\rightarrow L+1\left\rangle \right.-c.c.(0.6018)$\tabularnewline
 &  &  &  & $\mid H-1\rightarrow L+2\left\rangle \right.-c.c.(0.0854)$\tabularnewline
\hline 
 & $1^{3}A{}_{g}^{-}$  & 4.50  & 0.191  & $\mid H\rightarrow L;H-1\rightarrow L+1\left\rangle \right.(0.6134)$\tabularnewline
 &  &  &  & $\mid H\rightarrow L;H\rightarrow L+2\left\rangle \right.+c.c.(0.3260)$\tabularnewline
\hline 
II  & $2^{3}A{}_{g}^{-}$  & 4.85  & 0.715  & $\mid H\rightarrow L;H-1\rightarrow L+1\left\rangle \right.(0.5609)$\tabularnewline
 &  &  &  & $\mid H\rightarrow L+4\left\rangle \right.-c.c.(0.3636)$\tabularnewline
 & $2^{3}B_{1g}^{-}$  & 4.88  & 0.435  & $\mid H\rightarrow L+5\left\rangle \right.+c.c.(0.5435)$\tabularnewline
 &  &  &  & $\mid H-2\rightarrow L+1\left\rangle \right.-c.c.(0.2464)$\tabularnewline
\hline 
III  & $4^{3}B_{1g}^{-}$  & 5.17  & 1.776  & $\mid H\rightarrow L;H\rightarrow L+3\left\rangle \right.-c.c.(0.5398)$\tabularnewline
 &  &  &  & $\mid H-2\rightarrow L+1\left\rangle \right.-c.c.(0.1420)$\tabularnewline
\hline 
IV  & $5^{3}B_{1g}^{-}$  & 6.23  & 1.114  & $\mid H\rightarrow L;H\rightarrow L+3\left\rangle \right.-c.c.(0.5398)$\tabularnewline
 &  &  &  & $\mid H-2\rightarrow L+1\left\rangle \right.-c.c.(0.1420)$\tabularnewline
 & $5^{3}A{}_{g}^{-}$  & 6.26  & 0.139  & $\mid H\rightarrow L;H-1\rightarrow L+5\left\rangle \right.-c.c.(0.3769)$\tabularnewline
 &  &  &  & $\mid H\rightarrow L+1;H-2\rightarrow L+1\left\rangle \right.-c.c.(0.3675)$\tabularnewline
\hline 
V  & $10^{3}A{}_{g}^{-}$  & 6.77  & 0.159  & $\mid H\rightarrow L+2;H-1\rightarrow L+1\left\rangle \right.-c.c.(0.4111)$\tabularnewline
 &  &  &  & $\mid H\rightarrow L;H\rightarrow L+2\left\rangle \right.-c.c.(0.2945)$\tabularnewline
 & $11^{3}B_{1g}^{-}$  & 6.85  & 0.441  & $\mid H\rightarrow L;H\rightarrow L+7\left\rangle \right.+c.c.(0.2565)$\tabularnewline
 &  &  &  & $\mid H-2\rightarrow L+5\left\rangle \right.+c.c.(0.2555)$\tabularnewline
\hline 
\end{tabular}
\end{table}

\par\end{center}

\begin{center}
\begin{table}[H]
\protect\protect\protect\caption{\textcolor{black}{Properties of the excited states leading to various
peaks in the triplet absorption spectrum of heptacene computed using
the MRSDCI method, and the standard parameters in the PPP model Hamiltonian.
The rest of the information is same as in the caption of Table \ref{tab:acene2-pa-std}.\label{tab:acene7-pa-std}}}

\vspace{0.25cm}

\centering{}%
\begin{tabular}{|c|c|c|c|c|}
\hline 
Peak  & State  & E (eV)  & Transition  & Dominant Contributing Configurations\tabularnewline
 &  &  & Dipole (\AA )  & \tabularnewline
DF  & $1^{3}B_{1g}^{+}$  & 1.42  & 0  & $\mid H\rightarrow L+1\left\rangle \right.+c.c.(0.5809)$\tabularnewline
 &  &  &  & $\mid H\rightarrow L+4\left\rangle \right.+c.c.(0.1313)$\tabularnewline
DF  & $1^{3}A{}_{g}^{+}$  & 3.07  & 0  & $\mid H\rightarrow L;H-1\rightarrow L+1\left\rangle \right.(0.4476)$\tabularnewline
 &  &  &  & $\mid H\rightarrow L+5\left\rangle \right.-c.c.(0.2726)$\tabularnewline
\hline 
I  & $1^{3}B_{1g}^{-}$  & 3.31  & 2.411  & $\mid H\rightarrow L+1\left\rangle \right.-c.c.(0.5652)$\tabularnewline
 &  &  &  & $\mid H-1\rightarrow L+2\left\rangle \right.-c.c.(0.1612)$\tabularnewline
\hline 
II  & $1^{3}A{}_{g}^{-}$  & 4.45  & 0.438  & $\mid H\rightarrow L;H-1\rightarrow L+1\left\rangle \right.(0.6770)$\tabularnewline
 &  &  &  & $\mid H\rightarrow L+5\left\rangle \right.+c.c.(0.2778)$\tabularnewline
 & $2^{3}A{}_{g}^{-}$  & 4.58  & 0.627  & $\mid H\rightarrow L;H-1\rightarrow L+1\left\rangle \right.(0.5450)$\tabularnewline
 &  &  &  & $\mid H-1\rightarrow L+3\left\rangle \right.-c.c.(0.3565)$\tabularnewline
\hline 
III  & $3^{3}B_{1g}^{-}$  & 5.46  & 2.368  & $\mid H\rightarrow L;H\rightarrow L+3\left\rangle \right.-c.c.(0.5201)$\tabularnewline
 &  &  &  & $\mid H\rightarrow L+1;H-5\rightarrow L\left\rangle \right.c.c.(0.1864)$\tabularnewline
\hline 
IV  & $9^{3}A{}_{g}^{-}$  & 6.48  & 0.314  & $\mid H\rightarrow L+2;H-1\rightarrow L+1\left\rangle \right.-c.c.(0.3296)$\tabularnewline
 &  &  &  & $\mid H\rightarrow L;H-1\rightarrow L+4\left\rangle \right.-c.c.(0.2959)$\tabularnewline
 & $6^{3}B_{1g}^{-}$  & 6.51  & 1.142  & $\mid H\rightarrow L+1;H-5\rightarrow L\left\rangle \right.c.c.(0.4693)$\tabularnewline
 &  &  &  & $\mid H\rightarrow L+1;H-1\rightarrow L+3\left\rangle \right.c.c.(0.3070)$\tabularnewline
\hline 
\end{tabular}
\end{table}

\par\end{center}

\begin{center}
\begin{table}[H]
\protect\protect\protect\caption{\textcolor{black}{Properties of the excited states leading to various
peaks in the triplet absorption spectrum of heptacene computed using
the MRSDCI method, and the screened parameters in the PPP model Hamiltonian.
The rest of the information is same as in the caption of Table \ref{tab:acene2-pa-std}.
\label{tab:acene7-pa-scr}}}

\vspace{0.25cm}

\centering{}%
\begin{tabular}{|c|c|c|c|c|}
\hline 
Peak  & State  & E (eV)  & Transition  & Dominant Contributing Configurations\tabularnewline
 &  &  & Dipole (\AA )  & \tabularnewline
DF  & $1^{3}B_{1g}^{+}$  & 1.30  & 0.0  & $\mid H\rightarrow L+1\left\rangle \right.+c.c.(0.5728)$\tabularnewline
 &  &  &  & $\mid H-1\rightarrow L+2\left\rangle \right.+c.c.(0.1278)$\tabularnewline
DF  & $1^{3}A{}_{g}^{+}$  & 2.59  & 0.0  & $\mid H\rightarrow L;H-1\rightarrow L+1\left\rangle \right.(0.7028)$\tabularnewline
 &  &  &  & $\mid H\rightarrow L+2;H-1\rightarrow L+1\left\rangle \right.-c.c.(0.2562)$\tabularnewline
\hline 
I  & $1^{3}B_{1g}^{-}$  & 2.83  & 2.708  & $\mid H\rightarrow L+1\left\rangle \right.-c.c.(0.5973)$\tabularnewline
 &  &  &  & $\mid H-1\rightarrow L+2\left\rangle \right.-c.c.(0.0932)$\tabularnewline
\hline 
II  & $1^{3}A{}_{g}^{-}$  & 3.84  & 0.756  & $\mid H\rightarrow L;H-1\rightarrow L+1\left\rangle \right.(0.8354)$\tabularnewline
 &  &  &  & $\mid H\rightarrow L+2;H-1\rightarrow L+1\left\rangle \right.-c.c.(0.1082)$\tabularnewline
\hline 
III  & $2^{3}A{}_{g}^{-}$  & 4.28  & 0.157  & $\mid H\rightarrow L;H-1\rightarrow L+1\left\rangle \right.(0.7992)$\tabularnewline
 &  &  &  & $\mid H-1\rightarrow L+1;H-2\rightarrow L\left\rangle \right.c.c.(0.1401)$\tabularnewline
 & $2^{3}B_{1g}^{-}$  & 4.28  & 0.624  & $\mid H\rightarrow L+4\left\rangle \right.+c.c.(0.4639)$\tabularnewline
 &  &  &  & $\mid H-1\rightarrow L+2\left\rangle \right.-c.c.(3640)$\tabularnewline
\hline 
IV  & $3^{3}A{}_{g}^{-}$  & 4.49  & 0.614  & $\mid H\rightarrow L+5\left\rangle \right.+c.c.(0.5809)$\tabularnewline
 &  &  &  & $\mid H-1\rightarrow L;H-1\rightarrow L+1\left\rangle \right.(0.1796)$\tabularnewline
 & $3^{3}B_{1g}^{-}$  & 4.45  & 0.519  & $\mid H-1\rightarrow L+2\left\rangle \right.-c.c.(0.4166)$\tabularnewline
 &  &  &  & $\mid H\rightarrow L+4\left\rangle \right.+c.c.(3838)$\tabularnewline
\hline 
V  & $4^{3}B_{1g}^{-}$  & 4.8  & 2.177  & $\mid H\rightarrow L;H\rightarrow L+3\left\rangle \right.-c.c.(0.5665)$\tabularnewline
 &  &  &  & $\mid H\rightarrow L+1;H-1\rightarrow L+3\left\rangle \right.c.c.(0.1631)$\tabularnewline
\hline 
VI  & $5^{3}B_{1g}^{-}$  & 5.63  & 1.361  & $\mid H\rightarrow L+1;H-5\rightarrow L\left\rangle \right.-c.c.(0.5143)$\tabularnewline
 &  &  &  & $\mid H\rightarrow L+1;H-1\rightarrow L+3\left\rangle \right.-c.c.(0.2377)$\tabularnewline
\hline 
VII  & $8^{3}B_{1g}^{-}$  & 6.11  & 0.508  & $\mid H-2\rightarrow L+4\left\rangle \right.+c.c.(0.3988)$\tabularnewline
 &  &  &  & $\mid H\rightarrow L+1;H-5\rightarrow L\left\rangle \right.-c.c.(0.3343)$\tabularnewline
 & $11^{3}A{}_{g}^{-}$  & 6.17  & 0.299  & $\mid H\rightarrow L;H-1\rightarrow L+4\left\rangle \right.-c.c.(0.3075)$\tabularnewline
 &  &  &  & $\mid H\rightarrow L+2;H-1\rightarrow L+1\left\rangle \right.-c.c.(0.3065)$\tabularnewline
\hline 
\end{tabular}
\end{table}

\par\end{center}

 \bibliographystyle{apsrev4-1}
\bibliography{small_acene}

\end{document}